\documentclass[aps,pra,showpacs,superscriptaddress,twocolumn,10pt]{revtex4-1}
\usepackage{hyperref}
\usepackage[cm]{fullpage}
\usepackage{graphicx}
\usepackage[english]{babel}
\usepackage{amsmath}
\usepackage{amssymb}
\usepackage{cancel}
\usepackage{comment}
\usepackage{color}
\usepackage[normalem]{ulem}
\newcommand{\ket}[1]{{\left\vert {#1} \right\rangle}}
\newcommand{\bra}[1]{{\left\langle {#1} \right\vert}}

\newcommand{\eq}{Eq.~}
\newcommand{\eqs}{Eqs.~}
\newcommand{\fig}{Fig.~}

\newcommand{\cf} {cf.~}
\newcommand{\ug} {\!=\!}

\newcommand{\piu} {\!+\!}
\newcommand{\meno} {\!-\!}

\newcommand{\kket}[1]{\left.\left\vert#1\right\rangle\right\rangle}
\newcommand{\bbra}[1]{\left\langle\left\langle#1\right\vert\right.}

\makeatletter
	\renewcommand{\maketag@@@}[1]{\hbox{\m@th\normalsize\normalfont#1}}
\makeatother
\begin{document}
\title{Heat flux and quantum correlations in dissipative cascaded systems}
\author{Salvatore Lorenzo}
\affiliation{Dipartimento di Fisica e Chimica, Universit$\grave{a}$  degli Studi di Palermo, via Archirafi 36, I-90123 Palermo, Italy}   
\author{Alessandro Farace}
\affiliation{NEST, Scuola Normale Superiore and Istituto Nanoscienze-CNR, I-56126 Pisa, Italy} 
\author{Francesco Ciccarello}
\affiliation{NEST, Istituto Nanoscienze-CNR and Dipartimento di Fisica e Chimica, Universit$\grave{a}$  degli Studi di Palermo, via Archirafi 36, I-90123 Palermo, Italy}   
\author{G. Massimo Palma}
\affiliation{NEST, Istituto Nanoscienze-CNR and Dipartimento di Fisica e Chimica, Universit$\grave{a}$  degli Studi di Palermo, via Archirafi 36, I-90123 Palermo, Italy}   
\author{Vittorio Giovannetti}
\affiliation{NEST, Scuola Normale Superiore and Istituto Nanoscienze-CNR, I-56126 Pisa, Italy}     
\date{\today}
\begin{abstract}
We study the dynamics of heat flux in the thermalization process of a pair of identical quantum system that interact dissipatively with a reservoir in a {\it cascaded} fashion. Despite the open dynamics of the bipartite system $S$ is globally Lindbladian, one of the subsystems ``sees" the reservoir in a state modified by the interaction with the other subsystem and hence it undergoes a non-Markovian dynamics. As a consequence, the heat flow exhibits a  non-exponential time behaviour which can greatly deviate from the case where each party is independently coupled to the reservoir. We investigate both thermal and correlated initial states of $S$ and show that the presence of correlations at the beginning can considerably affect the heat flux rate. We carry out our study in two paradigmatic cases -- a pair of harmonic oscillators with a reservoir of bosonic modes and two qubits with a reservoir of fermionic modes  -- and compare the corresponding behaviours. In the case of qubits and for initial thermal states, we find that the trace distance discord is at any time interpretable as the correlated contribution to the total heat flux.
\end{abstract}
\maketitle
%
%
\section{Introduction}
\label{SecIntro}
 
A fundamental thermodynamic quantity is the amount of energy that can be extracted from non-equilibrium systems.
The field of quantum thermodynamics \cite{Mahler2004,QT1,QT2,QT3} is currently experiencing a considerable effort to understand the concepts of work and heat within quantum mechanics\cite{Mukamel2003,Talkner2007,Campisi2009,Esposito2009,Esposito2010,Campisi2011}. 
While work is commonly analyzed in the presence of external coherent control on the system \cite{QT2,QT3,Talkner2007}, heat is associated to energy changes that are due to some system-bath interaction \cite{QT1,Campisi2009,Mahler2008}.
Quantum Thermodynamics tackles heat transfer by modelling the  system-bath interactions as a quantum mechanical process mathematically described, under weak-coupling assumptions, by the Lindblad generator \cite{Lindblad1976}.
Scenarios featuring consecutive interactions between individual elements of a  quantum multipartite system and their own local environments
have recently been investigated \cite{Plenio2007,Caruso2010,Benenti2009,REVIEW} and the study of these correlated channels has made clear that 
interesting new features emerge in the presence of correlations.

Given the quantum mechanical nature of such processes, an interesting question is if, and how, the heat flux between a multipartite system and its reservoir can be affected by intra-system {\it quantum} correlations (QCs) which are present in the initial state. In particular, one can investigate whether QCs, either in the form of entanglement \cite{Horodecki2009a} or quantum discord~\cite{Modi2012a}, are fundamental resources for the heat transfer mechanism. Note that a similar issue was tackled in the completely different framework of quantum biology, see e.g.~\cite{BIO1,BIO2}.

It is straightforward to predict that, if the various subsystems are not directly coupled and the reservoir is sufficiently large to prevent any cross-talking, then
correlations do not play any role. In such cases, the heat flux emerging from a composite system is the same for all the initial states admitting the same local representation, regardless of the presence of correlations among its constituents. 
The scenario however changes drastically if we do introduce interactions among the various subsystems or if the reservoir ``sees" the compound systems as a unique object (so called common bath). For instance, it is well known that a strong coupling between two atoms 
can inhibit energy dissipation via the formation of dark states effectively decoupled from the reservoir~\cite{lambro-book}. 
In all these cases, quantum coherence (at the level of either initial correlations or interactions) plays a major role. 

In this paper, we shed light on such issues in the case of a cascade bipartite system where energy flows between its subsystems along a specific direction (say from subsystem 1 to subsystem 2 but not the opposite). 

Although thermal equilibrium with the heat bath is always reached after an infinite amount of time, a stronger or weaker heat flux can be obtained by engineering correlations in the initial state of the system, giving rise to very different timescales for the thermalization process. This means that the same amount of energy, stored into different configurations of the system, can be retrieved faster or slower according to the chosen state preparation.
In our study, we adopt the master equation approach developed by Gardiner {\it et al.}~\cite{Gardiner1985a,Gardiner1994a} in the case of bosonic baths and recently generalized by two of us~\cite{Giovannetti2012a} via a collision-model-based approach.  Within this framework, we discuss both the case of continuos-variable systems (two quantum harmonic oscillators) and the case of two-level systems (a pair of qubits \cite{nc}) showing how the presence of initial correlations can influence the system dynamics by speeding up or slowing down the energy flux to or from the reservoir. 
Interestingly enough, we find that in both scenarios, while entanglement among the subsystems appears not to play an essential role, the extremal performances in terms of heat flux rate take place in the presence of high values of non-classical correlations~\cite{Modi2012a} in the initial state of the system. Yet, strong quantum correlations are not sufficient to ensure faster or slower energy transfer. This is particularly true in the continuous-variable case where states featuring the maximum level of non-classicality do not show any difference in terms oh heat fluxes with respect to the completely uncorrelated case. 
While our analysis is of a conceptual nature (the systems under study being rather idealized) the  effects we describe  may find potential applications in designing  more efficient energy storage units or energy filters.   
 
The outline of the paper is as follows. In Section \ref{SecSys}, we describe the model under consideration and the master equation describing its open dynamics under a cascade interaction with the reservoir. In Section \ref{heat-fluxes}, we investigate the general form of the total and local heat fluxes and show that the former can be decomposed into three contributions, one of which reflects the interaction between the subsystems mediated by the reservoir.
In Section \ref{time_heat_fluxes}, we address the general time dependance of heat fluxes for both harmonic oscillators and qubits.
In Section \ref{HO}, (case of harmonic oscillators) and \ref{SubsecCorrHeat} (qubits) we analyze extensively the heat flux dynamics when the initial state of the open system is thermal or correlated (but locally thermal). We furthermore investigate on the role of initial QCs. In Section VII, we show that in some cases the correlated heat flux can be directly connected to a discord-like measure of QCs. Finally, in Section~\ref{Conclusions}, we draw our conclusions.
\section{Model and master equation}
\label{SecSys}
We consider a bipartite open system $S$, consisting of a pair of subsystems $S_1$ and $S_2$, and a thermal reservoir $R$ modeled as a large ensemble of identical ancillas all in the same initial thermal state. The $S$-$R$ interaction occurs in {\it cascade}~\cite{Gardiner1993a}. $S_1$ interacts with $R$ through a sequence of system-ancilla collisions under the usual Born-Markov approximation \cite{BreuerBook}. $S_2$, instead, interacts with $R$ {\it modified} by the previous interaction with $S_1$, see \fig1(a). No direct mutual coupling between $S_1$ and $S_2$ is present. Yet, $R$ mediates an indirect coupling between them.
\begin{figure}[htbp]
	\begin{center}
	\includegraphics[trim=0pt 0pt 0pt 0pt, clip, width=0.4\textwidth]{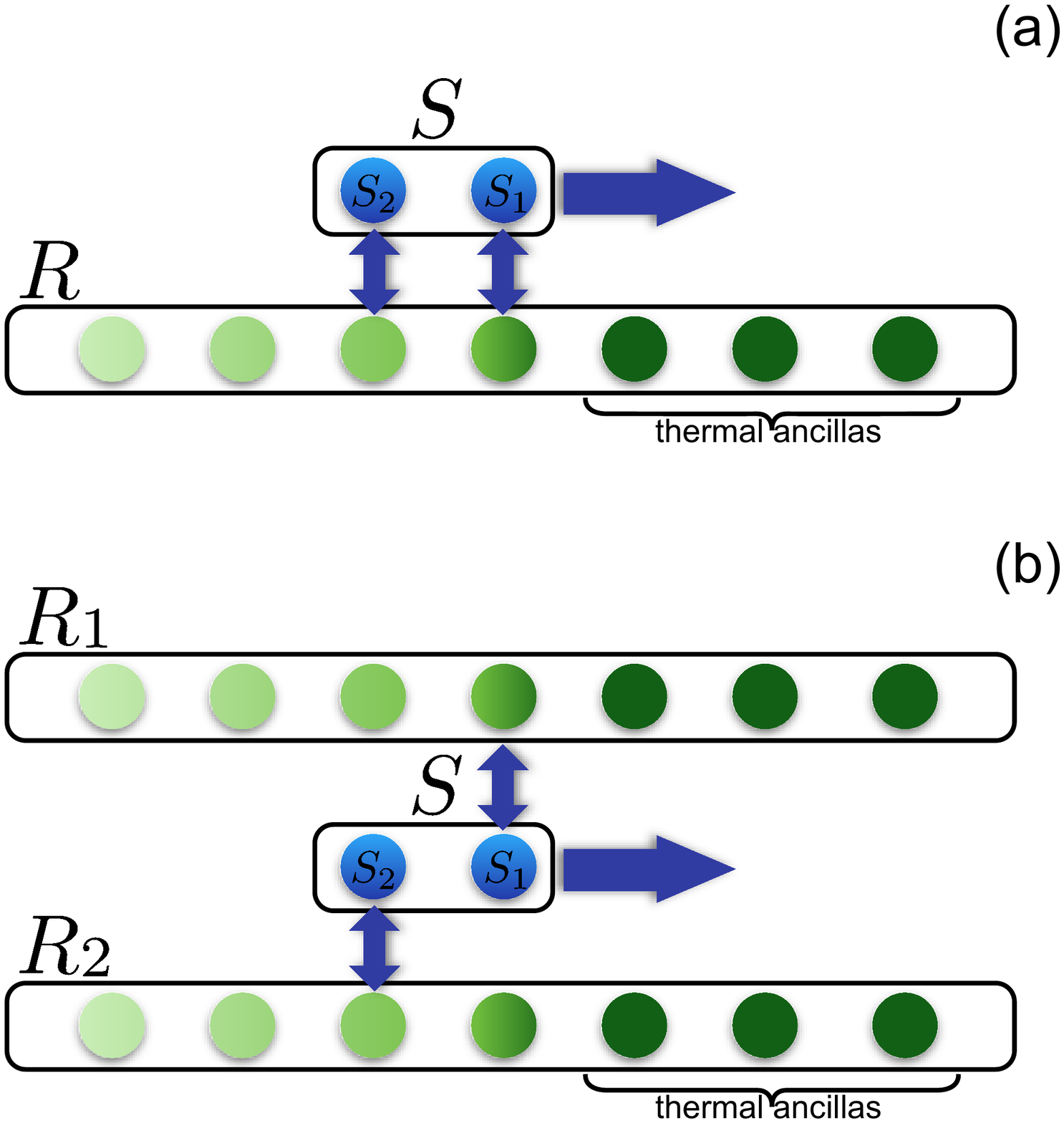}
	\caption{(Color online){(a) Sketch of the cascade interaction between $S$ and $R$. In this collision-model-based picture, $R$ is modelled as a large collection of ancillas. Each subpart $S_i$ of system $S$ interacts in succession with the reservoir ancillas. $S_1$ always interacts with ancillas that are in a thermal state. In contrast, $S_2$ encounters ancillas that have previously interacted with $S_1$ (hence they are no more in thermal state). For the sake of simplicity in our analysis the delay time which elapses between 
	the collision of a given ancilla elements with $S_1$ and the subsequent collision with $S_2$ is assumed to be negligible with respect to the other time scales of the system (see main text). (b) Sketch of the model where the cascade interaction has been removed. In this case the evolution of  $S_1$ and $S_2$ is the same as if they were interacting with two
	copies ($R_1$ and $R_2$) of the same reservoir.}}
	\label{FigSys}
	\end{center}
\end{figure}
Such indirect $S_1$-$S_2$ coupling is however {\it unidirectional}: $S_1$ affects the dynamics of $S_2$, but $S_2$ cannot influence $S_1$ in any way. 
The master equation in the $S$ state $\rho$ at time $t$ was derived long ago for bosonic baths through the input-output formalism \cite{Gardiner1985a,Gardiner1994a,Gardinerbook} and, quite recently, generalized to arbitrary baths by means of a collision-model-based approach~\cite{Giovannetti2012a}. 
To simplify the analysis, in what follows we shall assume that  the delay time between the $S_1$-$R$ and $S_2$-$R$ collisions is negligible compared to all the other system time scales. Still, the causal structure of the process holds: a collision between $S_1$ and a given ancilla of $R$ will anyway occur {\it before} the latter collides with $S_2$, see Fig.~\ref{FigSys}(a). 
Accordingly the master equation is of the Kossakowski-Lindblad form~\cite{BreuerBook} and reads
\begin{align}
	\dot \rho = & -\frac{i}{\hbar} [\hat{H}, \rho] +{\mathcal L}^{(\text{c})} (\rho)\;,
	\label{ME}
\end{align}
where $\hat{H} \ug\hat{H}_1\piu\hat H_2$ is the free Hamiltonian of $S$ ($\hat H_i$ is the local free Hamiltonian of the $i$th subsystem with $i\ug1,2$) while the cascade Lindbladian superoperator ${\mathcal L}^{(c)}$ is the sum of three terms according to
\begin{align}
{\mathcal L}^{(c)}= \mathcal{L}_1+ \mathcal{L}_2 + \mathcal{D}_{12}\,.\label{decomp}
 \end{align}
Here, $\mathcal{L}_i$ acts locally on $S_i$ only and coincides with the Lindblad superoperator that would be obtained if $S_i$ were in contact with $R$ in the absence of the other subsystem. The superoperator $\mathcal{D}_{12}$, instead, acts on both the subsystems and accounts for the cascade, i.e., one-way, $S_1\!\rightarrow\!S_2$ interaction mediated by $R$. The explicit forms of $\mathcal L_i$ and $\mathcal D_{12}$ will be given below in the cases of concern to this work (for simplicity, we will refer to such superoperators as ``dissipators" since we will focus on purely dissipative reservoirs). The general expressions for $\{\mathcal L_i\}$ and $\mathcal D_{12}$ can be found in \cite{Giovannetti2012a}.
For comparison, we will also analyze the case  where the cascade link is removed in a way that both systems interact with the reservoir $R$ independently, see Fig.~\ref{FigSys}(b).  Formally, this can be obtained by simply replacing in Eq.~(\ref{ME}) ${\mathcal L}^{(c)}$ with 
${\mathcal L}^{(ind)}\ug \mathcal{L}_1\piu \mathcal{L}_2$ (i.e., by setting $\mathcal{D}_{12}\ug0$).

We next illustrate the explicit form taken by $\mathcal L_i$ and $\mathcal D_{12}$ for a pair of CV variables (i.e., quantum harmonic oscillators) and qubits (i.e., two-level systems) in contact with a reservoir of harmonic oscillators and qubits, respectively.  
%
%
\subsection{Harmonic oscillators}
\label{SubsecOscill}
In this case, each subsystem $S_i$ is a quantum harmonic oscillator of frequency $\omega$ with associated bosonic annihilation and creation operators $\hat a_i$ and $\hat a^\dag_i$, respectively. The free Hamiltonian reads
\begin{align}
	\hat{H} =\hat H_1+\hat H_2= \hbar \omega (\hat a_1^\dagger \hat a_1 + \hat a_2^\dagger \hat a_2).
	\label{EqHamOscillator}
\end{align}
The reservoir $R$ consists of a large collection of bosonic modes. If the interaction Hamiltonian between the system and each reservoir mode does not feature counter-rotating terms (rotating-wave approximation), the local and non-local dissipators in Eq.\eqref{decomp} are then given by~\cite{Gardiner1994a,Giovannetti2012a}
\begin{align}
	\mathcal{L}_i(\rho) = & \,\tfrac{\gamma}{2}  (N+1) \left( 2 \hat a_i \rho \hat a_i^\dagger - \rho \hat a_i^\dagger \hat a_i - \hat a_i^\dagger \hat a_i \rho \right) \nonumber\\
	& + \tfrac{\gamma}{2}  N \left( 2 \hat a_i^\dagger \rho \hat a_i - \rho \hat a_i a_i^\dagger - \hat a_i \hat a_i^\dagger \rho \right), \label{Li1} \\
	\mathcal{D}_{12}(\rho) = &\, \gamma (N+1) \Big( \hat a_1 [\rho, \hat a_2^\dagger] + [\hat a_2, \rho] a_1^\dagger \Big) \nonumber\\
	& + \gamma N \Big( \hat a_1^\dagger [\rho, \hat a_2] + [\hat a_2^\dagger, \rho] \hat a_1 \Big).
	\label{D121}
\end{align}
Here, $\gamma$ coincides with the relaxation rate that would arise for each subsystem alone (assumed identical for the two subsystems), $N \ug1/(e^{\beta\hbar \omega}\meno1)$ is the thermal excitation number, $\beta\ug1/(k_{\rm B}T)$ is the inverse temperature, while $k_{\rm B}$ and $T$ are the Boltzmann constant and reservoir's temperature, respectively. 

%
\subsection{Qubits}
\label{SubsecQubits}
In this case, each subsystem $S_i$ is a two-level system (qubit) whose ground and excited states are $|g\rangle_i$ and $|e\rangle_i$, respectively. The corresponding energy gap is $\hbar\omega$. Let $\{\hat\sigma_{i\pm},\hat\sigma_{iz}\}$ be the usual pseudo-spin operators with $\hat \sigma_{i+}\ug\hat \sigma_{i-}^\dag\ug |e\rangle_i\langle g|$ and $\hat\sigma_{iz}\ug|e\rangle_i\langle e|\meno |g\rangle_i\langle g|$. The system's free Hamiltonian now reads
\begin{align}
	\hat{H} =\hat H_1+\hat H_2=  \frac{\hbar\omega}{2} \left( \hat\sigma_{1z}+\hat\sigma_{2z}\right).
	\label{H-qubits}
\end{align}

If the reservoir consists of a bath of qubits, under the rotating-wave approximation the local and non-local dissipators in Eq.\eqref{decomp} are given by \cite{Giovannetti2012a}
\begin{align}
\mathcal{L}_i=&\,\frac{\gamma}{4}(1{+}\xi)\left( 2 \hat\sigma_{i-}\rho \hat\sigma_{i+} \meno \rho \hat\sigma_{i+} \hat\sigma_{i-}\meno\hat\sigma_{i+} \hat\sigma_{i-}\rho \right)\nonumber\\
		     &+\frac{\gamma}{4}(1{-}\xi)\left( 2 \hat\sigma_{i+} \rho \hat\sigma_{i-}\meno\rho \hat\sigma_{i-} \hat\sigma_{i+} - \hat\sigma_{i-}\hat\sigma_{i+} \rho \right)\,,\label{Li2}\\
\mathcal{D}_{12}=&\,\frac{\gamma}{2}(1{+}\xi)\left(\hat\sigma_{1-}\left[\rho,\hat\sigma_{2+}\right]{+}\left[\hat\sigma_{2-} ,\rho\right]\hat\sigma_{1+}\right)\nonumber\\
		         &+\frac{\gamma}{2}(1{-}\xi)\left(\hat\sigma_{1+}\left[\rho,\hat\sigma_{2-} \right]{+}\left[\hat\sigma_{2+},\rho\right]\hat\sigma_{1-}\right)\,\label{D122}
\end{align}
with 
\begin{align}
\xi\ug\tanh\left[\frac{\hbar\omega/2}{k_{\rm B}T}\right]\label{xi}\,\,.
\end{align}
Note that \eqs(\ref{Li2}) and (\ref{D122}) have the same structure as \eqs(\ref{Li1}) and (\ref{D121}), but differ from these in the statistical nature of ladder operators (fermionic instead of bosonic) and the rates associated with the dissipators.

\section{Total and local heat fluxes} \label{heat-fluxes}

Both in the case of harmonic oscillators and qubits, any initial state $\rho(0)$ of the system asymptotically relaxes towards the stationary state
\begin{equation}
\rho(\infty)=\frac{e^{-\beta{\hat{H}_1}}}{{Z}}\otimes \frac{e^{-\beta{\hat{H}_2}}}{{Z}} \label{rho-infty}
\end{equation}
with ${Z} \ug{\rm Tr}_i[e^{-\beta{\hat{H}_i}}]$ (since the subsystems are identical, $Z$ does not depend on $i\ug1,2$). This can be checked by setting $\dot\rho\ug0$ in \eq\eqref{ME} and verifying that the resulting equation is fulfilled by state \eqref{rho-infty}, as proven in detail in Appendix \ref{AppA} for both harmonic oscillators and qubits. \eq\eqref{rho-infty} shows that the system thermalizes to the reservoir temperature. The asymptotic thermal state coincides  
with the one that would be obtained if $S_1$ and $S_2$ were in contact with $R$ independently [i.e., $\rho(\infty)$ is also the fixed point associated with the dissipator ${\cal L}^{(ind)}$]. Thereby, the presence of the correlated dissipator $\mathcal{D}_{12}$ in \eq(\ref{ME}) has no effect on the steady state, which is indeed fully factorized and does not feature any $S_1$-$S_2$ correlation, nor on the total amount of energy which is exchanged with the  reservoir, i.e., 
\begin{eqnarray}  
Q(\infty) &=&  \mbox{Tr}[ (\rho(\infty) - \rho(0)) \hat{H} ] \label{total}\;.
\end{eqnarray}
 However, significant correlations can in general arise during the {\it transient}. In turn, these correlations affect the way heat flows between $S$ -- specifically $S_2$ -- and $R$. The heat flux dynamics during such transient will be the focus of our analysis. 
 
 As in our model no external work is done on $S$, the {\it total} heat flux of $S$ -- we call it $J$ -- can be identified with the time derivative of the system energy $U\!=\!{\rm Tr}[\rho\hat{H}]$~\cite{Mahler2008}. Hence, at time $t$, the heat flux is calculated as $J(t)\ug\dot U\ug{\rm Tr}[\dot\rho(t)\hat{H}]$. 
In the case of the cascaded system, due to \eqs\eqref{ME} and \eqref{decomp}, this yields 
 \begin{align}
	{J}^{(c)}(t) =\mathcal{J}_{1}(t)+\mathcal{J}_{2}(t)+\mathcal{J}_{12}(t)
	\label{Jt}
\end{align}
with
\begin{eqnarray}
\mathcal J_i(t)&=&{\rm Tr}\left[\mathcal{L}_i\rho(t) \hat{H}\right]\!\equiv\!{\rm Tr}\left[\mathcal{L}_i\rho(t) \hat{H}_i\right]\!\,,\label{Ji}\\
\mathcal J_{12}(t)&=&{\rm Tr}\left[\mathcal{D}_{12}\rho(t) \hat{H}\right]\!\equiv\!{\rm Tr}\left[\mathcal{D}_{12}\rho(t) \hat{H}_2\right]\,.\label{J12}
\end{eqnarray}
\\
The total heat flux can thus be decomposed into three contributions, two of which stem from the local dissipators $\{\mathcal L_i\}$, one from the non-local dissipator $\mathcal D_{12}$. In \eqs(\ref{Ji}) and (\ref{J12}), the last identities show that $\hat H$ can be replaced by $\hat H_i$ ($\hat H_2$) in the calculation of $\mathcal J_i$ ($\mathcal D_{12}$). This is due to the identities
\begin{eqnarray}
{\rm Tr}[\mathcal L_1\rho\hat H_2]\ug{\rm Tr}[\mathcal L_2\rho\hat H_1]\ug{\rm Tr}[\mathcal D_{12}\rho\hat H_1]\ug0\,,\label{prop}
\end{eqnarray}
which can be straightforwardly proven upon use of \eqs(\ref{Li1}) and (\ref{Li2}) and the ciclic property of the trace.

As for the {\it local} heat fluxes of $S_1$ and $S_2$, by using \eqs(\ref{ME}), (\ref{decomp}), (\ref{Ji})-(\ref{prop}) these are respectively computed as
\begin{eqnarray}
J^{(c)}_1(t)&=&\dot{U}_1(t)\ug{\rm Tr}\left[\dot\rho(t) \hat{H}_1\right]\equiv\mathcal J_1(t),\label{J1}\\
J^{(c)}_2(t)&=&\dot{U}_2(t)\ug{\rm Tr}\left[\dot\rho(t) \hat{H}_2\right]\equiv\mathcal J_2(t)+\mathcal J_{12}(t)\,.\label{J2sum}
\end{eqnarray}
Upon comparison of these with the total heat flux (\ref{Jt}), we find $J^{(c)} (t)\ug J^{(c)}_1(t)\piu J^{(c)}_2(t)$ as expected. More importantly, the above equations show that, out of the three terms appearing in \eq(\ref{Jt}), $\mathcal J_1(t)$ accounts for the $S_1$ heat flux while the sum of the last two, i.e., $\mathcal J_2(t)\piu\mathcal J_{12}(t)$, is equal to $J^{(c)}_2(t)$. The correlated term $\mathcal J_{12}(t)$ therefore contributes only to the heat flux of $S_2$ (this is reasonable in light of the cascaded nature of the system dynamics). As anticipated, the reduced dynamics of $S_1$ fully coincides with that in the absence of $S_2$ since, upon trace over subsystem $S_2$ and using the cyclic property of the partial trace, \eq (\ref{ME}) yields $\dot\rho_1\ug\mathcal L_1\rho_1$. Correspondingly, $J^{(c)}_1(t)$ is just the same function as in the absence of $S_2$ since in \eq(\ref{J1}) $\rho(t)$ can be replaced with $\rho_1(t)$.

The heat flux associated with the identical and independent reservoirs model of  Fig.\ref{FigSys}(b) can be calculated in the same way. Again  the total flux is 
given by the sum of the fluxes  from $S_1$ and from $S_2$, i.e. $J^{(ind)} (t)\ug J^{(ind)}_1(t)\piu J^{(ind)}_2(t)$. Furthermore the heat flux $J^{(ind)}_1(t)$ from $S_1$ coincides with the one we computed for the cascaded system, i.e., $J^{(ind)}_1(t)=J^{(c)}(t) =\mathcal{J}_{1}(t)$, hence the two models give rise to the same reduced local dynamics for $S_1$. 
On the contrary 
 the heat flux from $S_2$, $J^{(ind)}_2(t)$  is rather different from $J^{(c)}_2(t)$. In particular, if we do assume that the initial state $\rho(0)$ is 
 locally indistinguishable for exchange of $S_1$ with  $S_2$, we have  $J^{(ind)}_2(t)=J^{(ind)}_1(t) = \mathcal{J}_{1}(t)$ (the local dissipative processes being identical). 
 Accordingly, we can write  
 \begin{eqnarray} 
J^{(ind)}(t) = 2  \mathcal{J}_{1}(t)\;, \label{fluxind} 
\end{eqnarray}
with $ \mathcal{J}_{1}(t)$ being the {\it same} function that appears on the right-hand-side of Eq.~(\ref{Jt}). 
It is finally worth stressing that due to the fact that both the cascade and the independent model yield the same total amount of dissipated energy~(\ref{total}) when integrated over the
whole evolution [i.e., $Q(\infty)= \int_0^{\infty} J^{(c)}(t) dt = \int_0^{\infty} J^{(ind)}(t) dt$], the following identity  holds
\begin{eqnarray} 
 \int_0^{\infty} [\mathcal{J}_{1}(t) -  \mathcal{J}_{2}(t)] dt = \int_0^{\infty} \mathcal{J}_{12}(t)dt\;. 
 \end{eqnarray}  

\section{Time dependance of heat fluxes}\label{time_heat_fluxes}
In this section, we show how the explicit procedure to calculate the three contributions to the total heat flux of \eq\eqref{decomp}, for harmonic oscillators and for qubits.

\subsection{Harmonic oscillators}
\label{time-harm}

In the case of harmonic oscillators, we focus on initial states $\rho(0)$ of $S$ that are {\it Gaussian} \cite{Ferraro2005a}. The linearity of the master equation \eqref{ME} alongside the assumption that the initial state of the ancillas of $R$ is thermal (hence Gaussian as well) ensures that the state of $S$ will remain Gaussian at any time $t$. To specify such states, let us introduce the position-momentum quadrature operators $\hat X_{j}{=}(\hat a_{j}^\dagger \piu\hat a_{j})/\!\sqrt{2}$ and $\hat Y_{j}{=}i (\hat a_{j}^\dagger \meno \hat a_{j})/\!\sqrt{2}$ with $j\ug1,2$. Correspondingly, let us define the four-dimensional vector operator $ \hat{\vec{ \chi} } \ug \{ \hat X_1,\hat Y_1,\hat X_2,\hat Y_2\}$.  By definition, a Gaussian state is fully specified by the expectation value of $\hat{\vec{ \chi} }$, i.e., $\{\langle\hat X_j\rangle, \langle\hat Y_j\rangle\}$, and by the covariance matrix $C_{mn} = \left< \frac{1}{2} ( \hat\chi_m  \hat\chi_n \piu  \hat\chi_n  \hat\chi_m) \right> - \left<  \hat\chi_m \right> \!\left< \hat \chi_n \right>$ with $m,n\ug1,...,4$. Throughout, we will consider states with vanishing first moments, i.e., $\langle\hat{\vec{ \chi} }(0) \rangle\ug0$, which amounts to assuming that the energy of $S$ is initially stored solely in the form of fluctuations. Indeed, correlations are entirely described by the fluctuations and our main concern is to highlight the interplay between heat fluxes and correlations. Each initial state we will consider, thereby, will be fully specified by the covariance matrix $C_{mn}$ (this has real entries).

For the class of initial states discussed so far, upon use of \eqs(\ref{Li1}), (\ref{D121}), (\ref{Ji}) and (\ref{J12}) the three heat fluxes on the right-hand side of \eq(\ref{Jt}) take the form
\begin{eqnarray}
	\mathcal{J}_1(t) &=& \hbar \omega \gamma  \left[\frac{C_{11}(t)  + C_{22}(t)}{2}- \left(N\piu\tfrac{1}{2} \right) \right]\,,\label{J1tcv}\\
\mathcal{J}_2(t) &=& \hbar \omega \gamma  \left[ \frac{C_{33}(t)  + C_{44}(t)}{2}- \left(N\piu\tfrac{1}{2} \right) \right]\,, \label{J2tcv}\\
\mathcal{J}_{1 2}(t) &=& \hbar \omega \gamma  \left[ C_{13}(t) + C_{24}(t) \right] \,.
	\label{J12tcv}
\end{eqnarray}
To calculate the explicit time evolution of the covariance matrix entries $C_{mn}(t)$ for a given initial state, it is convenient to use the Langevin equations \cite{Gardiner1994a} as illustrated in Appendix \ref{AppB} 

\subsection{Qubits}

In this case, with the help of \eqs(\ref{Li2}), (\ref{D122}), (\ref{Ji}) and (\ref{J12}) the contributions to the total heat flux on the right-hand side of \eq(\ref{Jt}) are calculated as
\begin{eqnarray}
\!\!\!\!\mathcal{J}_1(t)&\ug& \gamma  \Big[(1\! \piu\! \xi) [\rho_{11}(t)\piu \rho_{22}(t)]\!\meno\! (1\!\meno\!\xi) [\rho_{33}(t)\piu \rho_{44}(t)]\Big]\!,\,\,\,\,\,\,\,\,\,\,\label{J1tq}\\
\!\!\!\!\mathcal{J}_2(t)&\ug& \gamma  \Big[(1\! \piu\! \xi) [\rho_{11}(t)\piu \rho_{33}(t)]\!\meno\!(1\!\meno\!\xi) [\rho_{22}(t)\piu \rho_{44}(t)]\Big]\!,\,\,\,\,\,\,\,\,\,\,\label{J2tq}\\
\mathcal{J}_{12}(t)&\ug& 2 \gamma  \xi\,  \left[\rho_{23}(t)\piu \rho_{32}(t)\right]\,,\label{J12tq}
\end{eqnarray}
where $\rho_{mn}$, i.e., the matrix elements of $\rho$, are labeled according to the uncoupled basis of the $S$ Hilbert space $\{|ee\rangle_{12},|eg\rangle_{12},|ge\rangle_{12},|gg\rangle_{12}\}$. \eqs(\ref{J1tq})-(\ref{J12tq}) hold for an arbitrary initial two-qubit state $\rho(0)$.
To calculate the explicit time evolution of the density matrix entries $\rho_{mn}(t)$ for a given $\rho(0)$, it is convenient to use master equation \eqref{ME} in the Liouville space as shown in Appendix \ref{AppC}.

\section{Heat flux dynamics: harmonic oscillators}\label{HO}

In this section, we analyse the heat flux dynamics for a pair of harmonic oscillators. We will consider both thermal (hence uncorrelated) and correlated initial states of the reservoir. 

\subsection{Thermal initial states} \label{HO-uncorr}

\begin{figure}[htbp]
	\begin{center}
	\includegraphics[trim=0pt 0pt 0pt 0pt, clip, width=0.5\textwidth]{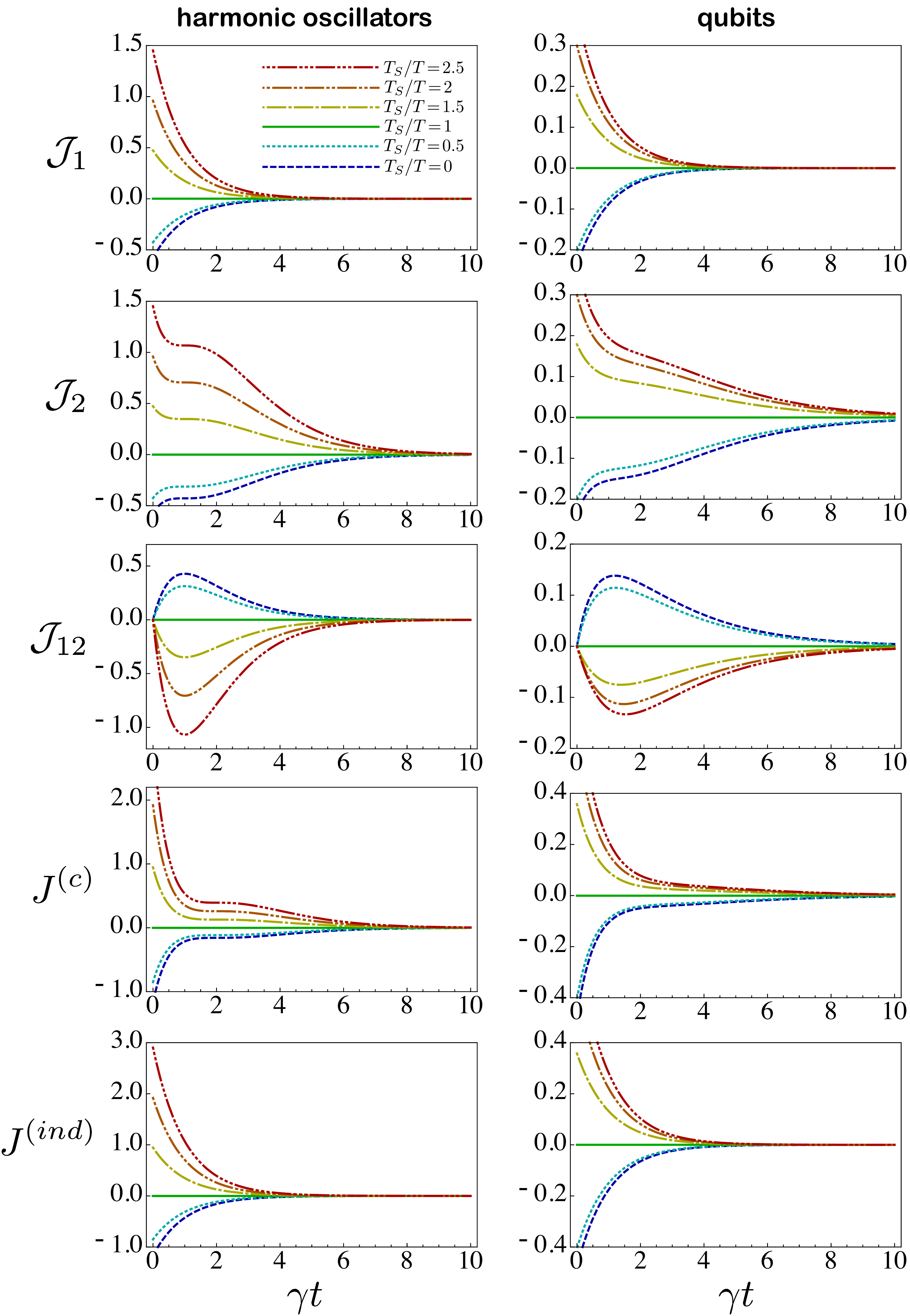}
	\caption{(Color online) Heat flows $\mathcal J_1$, $\mathcal J_2$, $\mathcal J_{12}$ and total heat flow $J^{(c)}$ against time in the case of harmonic oscillators (left-column plots) and qubits (right-column plots) for various temperatures $T_S$ (see the colour legend in the topmost left figure). As for the reservoir temperature, {{we have set it in such a way to have $k_{B}T/(\hbar \omega)\ug1$. Heat flows are expressed in  unit of $\hbar \omega \gamma$ and time is expressed in units of $\gamma^{-1}$}. In the bottom plots, we also report the behaviour of $J^{(ind)}$ for comparison.} {\color{blue}} }
	\label{heatflowsx2}
	\end{center}
\end{figure}

In this case, the pair of harmonic oscillators $S$ is initially in a thermal state $\rho(0)\ug e^{- \beta_S \hat{H}_1} \otimes e^{- \beta_S \hat{H}_2}/{Z_S}^2$, where $Z_S={\rm Tr}_i[e^{- \beta_S \hat{H}_i}]$, $\beta_S=1/(k_B T_S)$ and  $T_S$ is the system initial temperature. Note that, due to the lack of a direct coupling between $S_1$ and $S_2$, in such situation the two subsystems are initially fully {\it uncorrelated} and identical under mutual exchange.
Such initial conditions correspond to a covariance matrix whose only non-zero entries are ${C}_{ii}(0)\ug N_S\piu1/2$ for any $i\ug1,..,4$. Here, $N_S \ug1/(e^{\beta_S \hbar \omega}\meno1)$ is the initial average number of excitations in either $S$'s subsystem, which in general differs from $N$ (average number of excitations at the reservoir temperature). With the help of \eqs(\ref{J1tcv})-(\ref{J12tcv}) and Appendix \ref{AppB}, the explicit time dependances of $\mathcal J_1$, $\mathcal J_2$ and $\mathcal J_{12}$ is shown to be
\begin{eqnarray}
	\mathcal{J}_1(t)&=& \hbar \omega \gamma (N_S\meno N) e^{-\gamma t}\,,\label{J1tuncorr}\\
	\mathcal{J}_2(t) &\ug&(1\piu \gamma^2t^2)\mathcal{J}_1(t)\,,\,\mathcal{J}_{12}(t) \ug-2\gamma t\,\mathcal{J}_1(t)\label{J2-uncorr}\,,  
\end{eqnarray}
and hence the heat flux of $S_2$ [\cf\eq(\ref{J2sum})] for the cascade model reads
\begin{eqnarray}
	J^{(c)}_2(t) &\ug& (1-\gamma t)^2 \mathcal{J}_1(t) \,,\label{Jt-uncorr}
\end{eqnarray}
so that 
\begin{eqnarray}
	J^{(c)} (t) &\ug& [1+ (1-\gamma t)^2] \mathcal{J}_1(t)  \nonumber \\ &=&  \hbar \omega \gamma (N_S\meno N)\; [1+ (1-\gamma t)^2] e^{-\gamma t} \,,\label{Jtcasc}
\end{eqnarray}
In Fig.~\ref{heatflowsx2} (first column), we plot $J^{(c)}(t)$ and its the three components $\{\mathcal J_1(t), \mathcal J_2(t),\mathcal J_{12}(t)\}$ for different values of $T_S$ both above and below the reservoir's temperature $T$ which is chosen to be comparable with the typical energy scale of the system (specifically we assume $k_{B}T/( \hbar \omega) = 1$).
As expected, the heat flux of $S_1$ exponentially decays or increases [depending on the sign of ($N_S\meno N$)] at the rate $\gamma$. In contrast, both $\mathcal J_2(t)$ and $\mathcal J_{12}(t)$ exhibit non-exponential behaviour. The correlated heat $\mathcal J_{12}(t)$, in particular, has a non-monotonic behaviour: its absolute value grows from zero until it reaches a maximum at $\gamma t\ug1$ and then decreases. Also, note that the sign of $\mathcal J_{12}(t)$ is always opposite to that of $\mathcal J_1(t)$. 
The non-monotonic behaviour of $J^{(c)}_2(t)$ affects the total heat flow $J^{(c)}(t)$ to a significant extent. To better appreciate this 
consider  the scenario in which $S_1$ and $S_2$ are fully independent. The total flux in this case is expressed by Eq.~(\ref{fluxind}), i.e. 
\begin{eqnarray}
	J^{(ind)} (t) &\ug& 2  \mathcal{J}_1(t) =  2 \;\hbar \omega \gamma (N_S\meno N)\; e^{-\gamma t} \,.\label{Jtind}
\end{eqnarray}
By a direct comparison with Eq.~(\ref{Jtcasc}) it follows that the cascading mechanism makes $|J^{(c)}|$ lower (higher) than $|J^{(ind)}|$ for times shorter (larger) than $\gamma t\ug2$ (while maintaining the same sign in any case).
 In particular for  $T_S>T$ this implies that, when connected in cascade, $S_1$ and $S_2$ tend to retain energy for a longer time.

\subsection{Correlated initial states} \label{HO-corr}

Next, we investigate the effect of initial correlations between $S_1$ and $S_2$ on the heat flux dynamics. Specifically, we consider initial states $\rho(0)$ such that $\rho_1(0)\ug{\rm Tr}_2[\rho(0)]\ug  e^{- \beta_S \hat{H}_1}/Z_S$ and $\rho_2(0)\ug{\rm Tr}_1[\rho(0)]\ug  e^{- \beta_S \hat{H}_2}/Z_S$ but $\rho(0)\!\neq\!\rho_{1}(0)\otimes \rho_2(0)$. In other words, one such state is {\it locally} equivalent to a tensor product of thermal states at the same temperature $T_S$ (like those addressed in Subsection \ref{HO-uncorr}) but we allow $S_1$ and $S_2$ to initially share some correlations. For the sake of simplicity, we will focus on the case where the reservoir is at zero temperature, i.e., we set $N\ug0$ throughout. 

In line with Subsection \ref{HO-uncorr}, the requirement that the state is locally thermal at the uniform temperature $T_S$ (corresponding to the average excitation number $N_S$) yields that the diagonal entries of the initial-state covariance matrix are ${C}_{ii}(0)\ug N_S\piu1/2$ for any $i\ug1,..,4$. The energy is then given by $U\ug \tfrac{1}{2} \hbar \omega {\rm Tr}[C(0)]\ug2 \hbar \omega C_{11}(0)$. The remaining entries of $C(0)$ are set to zero except for $C_{13}(0) \ug C_{31}(0)$ and $C_{24}(0) \ug C_{42}(0)$ that can be non-null. This is because, at an arbitrary time $t$, the only off-diagonal entries which the heat fluxes depend on are $C_{13}(t)$ and $C_{24}(t)$ [\cf\eq(\ref{J12tcv})]. Moreover, as shown by \eqs (\ref{C13}) and (\ref{C24}) in Appendix \ref{AppB}, the initial values of the remaining off-diagonal elements do not affect the heat-flux dynamics since these are fully decoupled from $\{C_{13}(t),\,C_{24}(t)\}$. To summarise, we study initial states having the form
\begin{align}
	C(0)=
	\begin{pmatrix}
		C_{11}(0) & 0 & C_{13}(0) & 0 \\
		0 & C_{11}(0) & 0 & C_{24}(0) \\
		C_{13}(0) & 0 & C_{11}(0) & 0 \\
		0 & C_{24}(0) & 0 & C_{11}(0)
	\end{pmatrix}\,.
	\label{CV-family}
\end{align}
A rigorous parametrization of the family of covariance matrices of the form (\ref{CV-family}) is presented in Appendix \ref{AppD}.

Clearly, the heat flux of $S_1$ is again given by \eq(\ref{J1tuncorr}) with $N\ug0$. This immediately implies that the total flux  $J^{(ind)} (t)$ for the independent system model  remains identical to  the one computed in Eq.~(\ref{Jtind}), and will not depend upon the presence of initial correlations.
On the contrary  with the help of \eqs(\ref{J1tcv})-(\ref{J12tcv}) and Appendix \ref{AppB} the two contributions to the $S_2$  heat flux for the cascade system are calculated as
\begin{eqnarray}
	\mathcal{J}_2(t) &\ug&(1\piu \gamma^2t^2)\mathcal{J}_1(t)\meno\hbar\omega\gamma t[C_{13}(0)\piu C_{24}(0)]e^{-\gamma t}\,,\,\,\,\,\,\,\label{J2-corr}\\  
	\mathcal{J}_{12}(t) &\ug&-2\gamma t\,\mathcal{J}_1(t)\piu\hbar\omega\gamma[C_{13}(0)\piu C_{24}(0)]e^{-\gamma t}\,.\label{J12-corr}
\end{eqnarray}
Upon sum of these we thus obtain 
\begin{eqnarray}
	J^{(c)}_2(t) &\ug& (1-\gamma t)^2 \mathcal{J}_1(t) \piu\hbar\omega\gamma(1\meno\gamma t)[C_{13}(0)\piu C_{24}(0)]e^{-\gamma t}. \nonumber \\
	\label{J22-corr}
\end{eqnarray}
\eqs(\ref{J2-corr})-(\ref{J22-corr}) generalize \eqs(\ref{J2-uncorr})-(\ref{Jt-uncorr}), featuring additional terms proportional to $C_{13}(0)\piu C_{24}(0)$. Importantly, the fact that the heat flux depends on such off-diagonal entries only through their {\it sum} entails that for states such  that $C_{13}(0)\ug- C_{24}(0)$, irrespective of $|C_{13}(0)|$, the presence of initial correlations has no effect on the heat flux dynamics.

To illustrate the typical behavior of the total heat flux in the general case, in figure 3(a) we plot the total flux 
$J^{(c)}(t)$ of Eq.~(\ref{Jt}) for $N_S\ug 1$ and $C_{13}(0)\ug C_{24}(0)\ug-0.7 N_S, 0, 0.7 N_S$. We point out that, as explained in Appendix \ref{AppD}, focusing on states such that $C_{13}(0)\ug C_{24}(0)$ does not cause loss of generality.
As shown by the plots, in contrast to figure 2, a major consequence of the presence of initial correlations is the non-monotonicity of the heat flux time. This can be proven in detail through a study of the derivative of $J^{(c)}(t)$, as resulting from the sum of \eqs(\ref{J1tuncorr}) and (\ref{J22-corr}). 
\begin{figure}[h!]
	\begin{center}
	\includegraphics[trim=0pt 0pt 0pt 0pt, clip, width=0.45\textwidth]{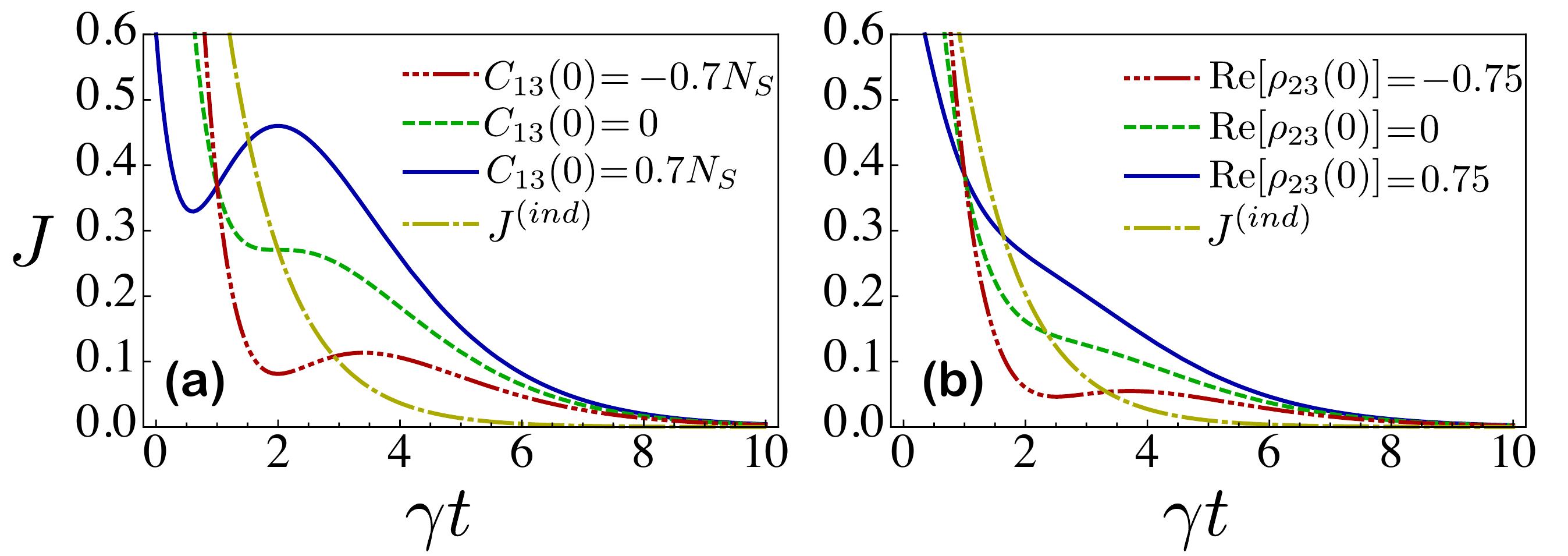}
	\caption{(Color online) (a): Time evolution of the total heat flux $J^{(c)}(t)\ug J^{(c)}_1(t)\piu J^{(c)}_2(t)$ for the cascade model in the case of harmonic oscillators for different choices of $C_{13}(0)\ug C_{24}(0)$, where we have set $N_S\ug1$ and $N\ug0$. (b): Time evolution of the total heat flux in the case of qubits for different choices of ${\rm Re}[\rho_{23}(0)]$, where we have set $\xi_S\ug0.25$ and $\xi\ug1$. In both cases, heat fluxes are in units of $\hbar \omega \gamma$ and time is in units of $\gamma^{-1}$ For comparison, the behaviour of $J^{ind}$ is also reported, which is independent of $C_{13}$ (${\rm Re}[\rho_{23}(0)]$) for harmonic oscillators (qubits).
	{\color{blue}{}}}
	\label{FigFlux}
	\end{center}
\end{figure}

The derivative reads
\begin{eqnarray}
\dot J^{(c)}(t)\ug\hbar\omega\gamma^2\! \left\{-(\gamma^2 N_S)t^2+\{\gamma  [C_{13}(0) \piu  C_{24}(0) \piu  4 N_S]\}t\right.\nonumber\\
\left.-2  [C_{13}(0) \piu  C_{24}(0) \piu  2 N_S] \right\}e^{-\gamma t}\nonumber\,.
\end{eqnarray}
As shown in Appendix \ref{AppD}, $|C_{13}(0) \piu  C_{24}(0)|\!\le\!2N_S$. Hence, in the above equation, the concave-down parabolic time function between curly brackets is non-positive at $t\ug0$. Moreover, this function has the two positive real roots
\begin{align}
	&t_{1} \ug \frac{2}{\gamma},\,\,\,\,\,\,\,\,\,t_{2} \ug \frac{2}{\gamma}\left[1+\frac{ C_{13}(0)\piu C_{24}(0)}{ 2N_S}\right]. \label{t1t2}
\end{align}
Thereby, $J^{(c)}(t)$ always exhibits a local minimum followed by a local maximum. Specifically, if $\left[C_{13}(0) \piu  C_{24}(0)\right]\!\le\!0$ the minimum occurs at $t_2$ and the maximum at $t_1\! >\! t_2$. Conversely, if $\left[C_{13}(0) \piu  C_{24}(0)\right]\!>\!0$ the minimum occurs at $t_1$ and the maximum at $t_2\! >\! t_1$.
 Such stationary points merge into a single inflection point, thus giving rise to a monotonic $J^{(c)}(t)$, for $C_{13}(0)\piu C_{24}(0)\ug0$.

Remarkably, not only the magnitude but even the sign of $C_{13}(0)\piu C_{24}(0)$ affects the heat flux in a significant way. This can be appreciated in figure 3(a),  which shows that the energy flow of $S$ into the reservoir proceeds slower when $C_{13}(0)\piu C_{24}(0)\!<\!0$. When the sum is positive, in contrast, most of the energy is released in the early stages of the dynamics. Such different behaviours can be better understood by calculating the value of $J^{(c)}(t)$ at $t=0$ and at times $t_{1,2}$ given by \eqref{t1t2}, which yields
\begin{eqnarray}
J^{(c)}(0)&\ug&\hbar\omega \gamma[2 N_S\piu C_{13}(0) \piu  C_{24}(0)],\\
J^{(c)}(t_1)&\ug& \hbar \omega \gamma \!\left\{2N_S \meno\left[{C_{13}(0) \piu  C_{24}(0)}\right] \right\} \!e^{-2},\label{Jt1}\\
J^{(c)}(t_2)&\ug& \hbar \omega \gamma  \!\left[2N_S \piu{C_{13}(0) \piu  C_{24}(0)} \right] \!e^{-\tfrac{[2N_S \piu C_{13}(0) \piu  C_{24}(0)]}{N_S}}\;. \nonumber \\\label{Jt2}
\end{eqnarray}

\noindent Hence, if $C_{13}(0) \piu  C_{24}(0)$ is positive, the first minimum always occurs at time $t_1$ and equals $J^{(c)}(t_1)$. As $e^{-2}\!\simeq\! 0.135$ [\cf\eq(\ref{Jt1})], in this case a drop of the heat flux of at least $\simeq\!86\%$ takes place after a time $2/\gamma$. The following rise of $J^{(c)}(t)$ is modest given that also the local maximum $J^{(c)}(t_2)$ is at most $\simeq\!14\%$ of the initial heat flux. Quite differently, if $C_{13}(0) \piu  C_{24}(0)$ is negative, the minimum occurs at time $t_2$, hence the corresponding drop amounts to the exponential factor in \eq \eqref{Jt2} which does not exceed $\simeq\!86\%$, this bound occurring  in the limiting case of very small $C_{13}(0) \piu  C_{24}(0)$. As this grows, the exponential factor rapidly approaches 1 (correspondingly the drop becomes less and less significant).

To characterise the release time of the system energy in more quantitative terms, in figure 4(a) we analyze $\gamma\tau_p$, namely the time (in units of $\gamma^{-1}$) taken by a certain percentage $p\%$ of the initial energy of $S$ to be lost into the reservoir. That is, we compute the energy lost up to some time $t$ as $Q^{(c)}(t) \equiv \int_0^t J^{(c)}(t') dt'$ and we search for the time $\tau_p$ at which $Q^{(c)}(\tau_p) = p\% \; Q^{(c)}(\infty)$ (i.e., $p\%$ of the total transferred energy). In figure 4(a), we plot $\gamma\tau_p$ versus $C_{13}(0)\piu C_{24}(0)$ for different values of the percentage $p$ (the outcomes are independent of $N_S$). The plots show that positive (negative) values of $C_{13}(0)+C_{24}(0)$ always speed up (slow down) the energy release compared to the uncorrelated case.  
\begin{figure}[htbp]
	\begin{center}
	\includegraphics[trim=0pt 0pt 0pt 0pt, clip, width=0.35\textwidth]{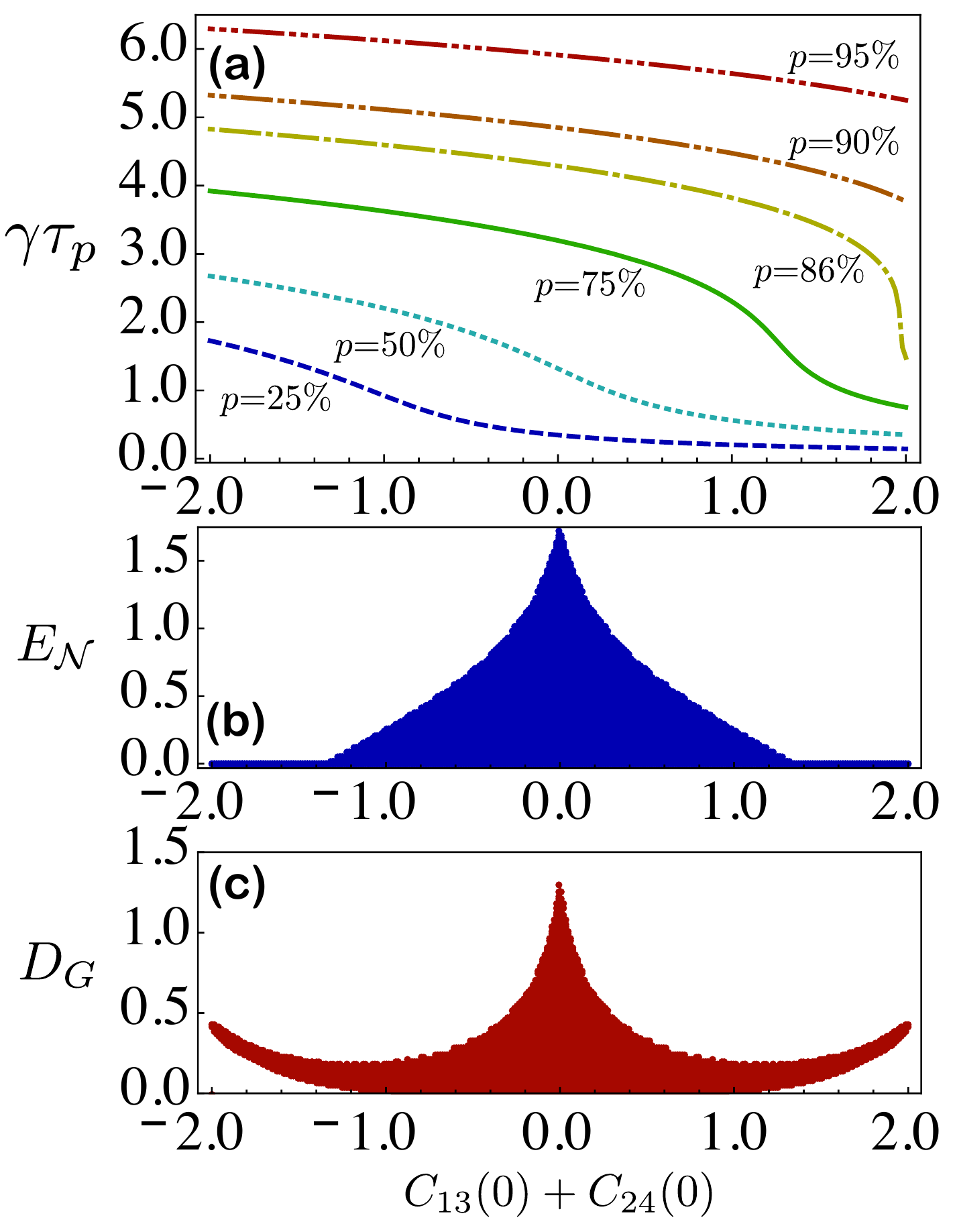}
	\caption{(Color online) (a): $\gamma \tau_p$ against $C_{13}(0)+C_{24}(0)$ for $p\ug95$ (red three-dotted-dashed line), $p \ug 90$ (orange two-dotted-dashed), $p\ug86$ (yellow dot-dashed), $p\ug75$ (green solid), $p\ug50$ (cyan dotted) and $p\ug25$ (blue dashed). (b): Entanglement, as measured by the logarithmic negativity $E_{\mathcal N}$, for all the states having the same value of $C_{13}(0)\piu C_{24}(0)$ as a function of $C_{13}(0)\piu C_{24}(0)$. (c): Gaussian discord $D_G$ for all the states having the same value of $C_{13}(0)\piu C_{24}(0)$ as a function of $C_{13}(0)\piu C_{24}(0)$. Throughout, we have set  $N_S{=}1$ and $N\ug0$.}
	\label{FigTime}
	\end{center}
\end{figure}

\subsection{Influence of initial quantum correlations} \label{HO-ent}

Next, we investigate the role played by typical measures of initial quantum correlations possessed by a state of the form (\ref{CV-family}). Traditionally, QCs have been associated with {\it entanglement} \cite{Horodecki2009a}. More recently, however, a new paradigm of QCs -- associated with the so called {\it quantum discord} -- has been put forward \cite{Modi2012a}. The need for introducing such a new type of QCs relies on the observation that, although separable, some bipartite states can feature correlations that are incompatible with classical physics. Specifically, here we will use logarithmic negativity \cite{Vidal2002a} ($E_\mathcal{N}$) and Gaussian discord \cite{Adesso2010a} ($D_G$) in order to quantify  entanglement and discord-like QCs, respectively. Details on both measures can be found in Appendix \ref{AppE}. Figures 5 shows density plots of logarithmic negativity (a) and Gaussian discord (b) on the $C_{13}(0)-C_{24}(0)$ plane for $N_S\ug1$ and $N\ug0$ (i.e., the paradigmatic instance addressed in the previous subsection).
Entanglement $E_{\mathcal {N}}$ arises only in two small regions next to the points $C_{13}(0)\ug -C_{24}(0)\ug \sqrt{N_S(N_S\piu 1)}$ and $C_{13}(0)\ug -C_{24}(0)\ug -\sqrt{N_S(N_S\piu 1)}$ \cite{footnote-cv}. In both cases, the corresponding state is close to an EPR state~\cite{Einstein1935a}. Instead, Gaussian discord $D_G$ is zero only at the point $C_{13}(0)\ug C_{24}(0)\ug 0$, which corresponds to a fully uncorrelated product state. It grows when the distance from this point increases. The steepest-increase directions are given by $C_{13}(0)\ug-C_{24}(0)$ (where also $E_{\mathcal {N}}$ increases) and $C_{13}(0)\ug C_{24}(0)$ (where instead entanglement is fully absent). 

\begin{figure*}[ht!]
		\includegraphics[trim=0pt 0pt 0pt 0pt, clip, width=0.8\textwidth]{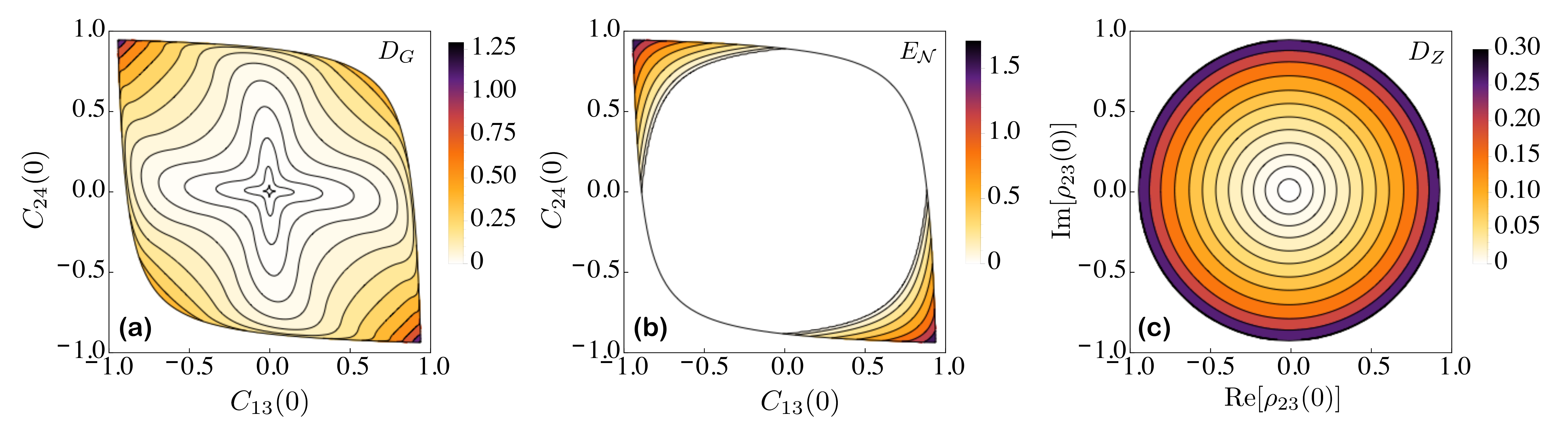}
	\caption{(Color online) Gaussian discord $D_G$ (a) and logarithmic negativity $E_\mathcal{N}$ (b) of a state (\ref{CV-family}) as functions of $C_{13}(0)$ and $C_{24}(0)$ for $N_S\ug1$ and $N\ug0$. (c): Quantum discord $D_Z$ of a state (\ref{rhoINI}) as a function of ${\rm Re}[\rho_{23}(0)]$ and ${\rm Im}[\rho_{23}(0)]$ for $\xi_{S}\ug0.25$ and $\xi\ug1$. The states (\ref{rhoINI}) considered for two qubits are never entangled.}
	\label{FigCorr}
\end{figure*}

As discussed in the previous subsection (see also Appendix \ref{AppD}), for any possible choice of $C_{13}(0)\ug C_{24}(0)\ug c_0$ there is a class of equivalent states (identified by $C_{13}(0) \piu C_{24}(0)\ug 2 c_0$) which exhibit the same heat flux dynamics [\cf\eqs(\ref{J2-corr}) and (\ref{J12-corr})]. The union of these classes coincides with the whole set of physical initial states. As shown in figure 5, all the states in a given class feature non-null $D_G$ [except for $C_{13}(0)\ug C_{24}(0)\ug0$], while a relevant fraction of them not entangled. In figures 4(b) and (c), for each value of $C_{13}(0) \piu C_{24}(0)$, we report all the possible values of $E_{\mathcal N}$ and $D_G$ in the corresponding equivalence class. We see that the states giving rise to the fastest and slowest energy release [corresponding to the highest and lowest values of $C_{13}(0)\ug C_{24}(0)\ug c_0$, respectively] are discordant but not entangled. For such states, Gaussian discord lies within a very narrow range (in general, the faster or slower the energy release the narrower the interval of possible values of $D_G$). Yet, based on figures 4 and 5, one can see that a high amount of discord does not necessarily lead to a fast or slow dissipation rate. Moreover, note that the most discordant state gives rise to the same heat flux time evolution as the completely uncorrelated state [see figure 4(c)]. The connection with energy release appears even weaker for entanglement as witnessed by the fact that, for each entangled state, there is always a separable one yielding the same heat flux dynamics [see figure 4(b)].  

Overall, the above analysis indicates that it is the peculiar {\it structure} of correlations -- instead of the featured amount of ``quantumness" -- that affects the heat flux dynamics. In particular, the quadratures that are most correlated plays the major role.
The optimal situation indeed occurs when the pairs $\{\hat X_1,\hat X_2\}$ and $\{\hat Y_1,\hat Y_2\}$ are equally (anti)correlated by the highest possible amount.
%
%
%
\section{Heat flux dynamics: qubits} \label{SubsecCorrHeat}

\subsection{Thermal initial states}\label{thermal-qubits}

$S$ now consists of a pair of qubits and both subsystems are initially in a local thermal state at temperature $T_S$, giving a joint (uncorrelated) initial state $\rho(0)= \exp[{-{\hat{H}_1}}/(k_{\rm B}T_S)]\exp[{-{\hat{H}_2}}/(k_{\rm B}T_S)]/Z_S^2$. 
 
The corresponding density matrix has zero off-diagonal entries, while the diagonal ones read
\begin{eqnarray}
\rho_{11}(0)&\ug&\frac{ (1\meno \xi_{S})^2}{4},\,\,\rho_{44}(0)\ug\frac{ (1\piu \xi_{S})^2}{4}\,,\label{diag1}\\
\rho_{22}(0)&\ug&\rho_{33}(0)\ug\frac{ 1\meno \xi_{S}^2}{4}\label{diag2}\,,
\end{eqnarray}
where $\xi_S$ is the value taken by \eq(\ref{xi}) for $T\ug T_S$.

One can use these (see Appendix \ref{AppC}) to calculate the time evolution of the density matrix elements entering \eqs\eqref{J1tq}-\eqref{J12tq}, hence the heat fluxes $\mathcal J_{1}(t)$, $\mathcal J_{2}(t)$, $\mathcal J_{12}(t)$ and the total heat flux $J^{(c)}(t)$. Unfortunately, the resulting analytic expressions are rather involved and uninformative (even in limiting cases). It turns out that no general exact relations as simple as those in \eqs \eqref{J2-uncorr} and (\ref{Jt-uncorr}) can be established. Yet, many of the salient features of the heat flux dynamics are qualitatively quite similar to those emerging for harmonic oscillators. This is shown by the right-column plots of figure 2, where we plot $\mathcal J_{1}$, $\mathcal J_{2}$, $\mathcal J_{12}$ and $J^{(c)}$ against time for different values of $T/T_S$ (the same considered in Section \ref{HO-uncorr}). The shape of each curve is quite similar to the corresponding one in the case of harmonic oscillators [a minor difference is that at intermediate times $\mathcal J_2(t)$ and $J^{(c)}(t)$ are not as flat as those for continuous-variable systems]. As a distinctive feature, though,  saturation appears at growing temperatures for each plotted quantity, which is clearly due to the fermionic nature of each subsystem as well as each reservoir mode.

\subsection{Correlated initial states}

In order to select a suitable family of correlated initial states $\rho(0)$, in full analogy with Subsection \ref{HO-corr}, we first require the local reduced qubit state to be locally thermal at temperature $T_S$. This entails that the only possible non-zero off-diagonal entries of $\rho(0)$ are $\rho_{23}(0)\ug\rho_{32}(0)^*$ and $\rho_{14}(0)\ug\rho_{41}(0)^*$ [the presence of extra off-diagonal entries would be incompatible with the constraint that each reduced state ${\rm Tr}_i\rho(0)$ has a diagonal form]. In a way similar to Subsection \ref{HO-corr}, to simplify the analysis, we further restrict to states such that $\rho_{14}(0)\ug\rho_{41}^*(0)\ug0$. Indeed, the heat fluxes in \eqs(\ref{J1tq})-(\ref{J12tq}) depend only on $\rho_{23}(t)$ and its c.c., which in turn are independent of $\rho_{14}(0)$ as shown in Appendix \ref{AppC}.

Therefore, 
\begin{equation}
\rho(0)=\frac{1}{4}\left(
\begin{smallmatrix}
(1-\xi_{S})^2& 0 & 0 & 0\\
0  & 1-\xi_{S}^2 &  \rho_{23}(0) &  0 \\
0  &  \rho_{23}(0)^*  & 1-\xi_{S}^2 &  0 \\
0  &  0  & 0 & (1+\xi_{S})^2
\end{smallmatrix}
\right)\,.
\label{rhoINI}
\end{equation}
The allowed values of $\rho_{23}(0)$ must fulfill the constraint 
\begin{equation}
|\rho_{23}(0)|\!\le1\meno\xi_{S}^2\label{qubitdomain}
\end{equation}
which follows from the requirement that density matrix (\ref{rhoINI}) be positive.

As in Subsection \ref{HO-corr}, we focus on the case of a zero-temperature reservoir (hence $\xi_{N}\ug1$). From \eqs(\ref{J1tq})-(\ref{J12tq}) and initial state (\ref{rhoINI}) -- see also Appendix \ref{AppC} -- the heat fluxes are calculated as
\begin{eqnarray}
\mathcal{J}_1(t)&\ug&\gamma  (1\meno \xi_{S} )e^{\meno \gamma t}\,,\nonumber\\
\mathcal{J}_2(t)&\ug&\gamma  \left\{\left(1\piu \gamma^2 t^2\right)(1\meno \xi_{S}) \piu 2 \left(1\meno \gamma t\meno e^{\meno \gamma t}\right)(1\meno \xi_{S})^2 \right.\nonumber\\
&&\left.\,\,\,\,\,\,\,\,\,\,\,\,\,\,\,\,\,\,\,\,\,\,\,\,\,\,\,\,\,\,\,\,\,\,\,\,- \gamma {\rm Re}[\rho_{23}(0)] t\right\}e^{\meno \gamma t}\,,\nonumber\\
\mathcal{J}_{12}(t)&\ug&\gamma \left\{2\left(1\meno e^{\meno \gamma t}\right) (1\meno \xi_{S})^2 \meno 2 \gamma t(1\meno \xi_{S})\piu {\rm Re}[\rho_{23}(0)]\right\}e^{\meno \gamma t}\,\nonumber.
\end{eqnarray}
Note that heat fluxes depend on the initial correlations through ${\rm Re}[\rho_{23}(0)]$. In figure 3(b), we use these results to plot the total heat flux, as given by \eq(\ref{Jt}), versus time for $\xi_{S}\ug0.25$ and three representative values of ${\rm Re}[\rho_{23}(0)]$.

As in the case of initial thermal states (see previous subsection), again we find a behaviour that qualitative resembles the one observed for harmonic oscillators (a minor difference occurs for the ${\rm Re}[\rho_{23}(0)]\ug0.75$ plot which does not feature stationary points but only concavity changes as time grows). This results from a comparison between figures 3(a) and 3(b), which shows that ${\rm Re}[\rho_{23}(0)]$ here behaves similarly to the parameter $C_{13}(0)\piu C_{24}(0)$ for harmonic oscillators. Negative (positive) values of ${\rm Re}[\rho_{23}(0)]$ cause a slow (fast) energy release.

In analogy with figure 4(a), in figure 6(a) we plot $\gamma\tau_p$ (time required to dissipate $p$\% of the initial energy) for $\xi_{S}\ug0$. The plots show that positive (negative) values of ${\rm Re}[\rho_{23}(0)]$ always speed up (slow down) the energy release compared to the uncorrelated case.
The relationship between the heat flux behaviour and the initial correlations can be better understood (see Appendix E) by expressing the superoperators (\ref{Li2}, \ref{D122}) and the initial state (\ref{rhoINI}) in the collective basis $ \lbrace |ee\rangle,|\Psi^+\rangle,|\Psi^-\rangle,|gg\rangle \rbrace$, where $|\Psi^\pm\rangle\!\equiv\!1/\sqrt{2}(|eg\rangle_{12}\pm|ge\rangle_{12})$. 
Such rearrangement shows that states $|\Psi^+\rangle$ and $|\Psi^-\rangle$ are coupled to the environment with different strengths. In particular, the singlet $|\Psi^-\rangle$ is fully decoupled from the environment for $T\ug0$. A positive initial value of $\rho_{23}$ means a smaller initial population of $|\Psi^-\rangle$ and therefore a faster energy release. A negative initial value of ${\rm Re}[\rho_{23}]$ means a larger initial population of $|\Psi^-\rangle$, hence a slower energy flow. This is shown in more detail in Appendix \ref{AppX}
\begin{figure}[htbp]
	\begin{center}
	\includegraphics[trim=0pt 0pt 0pt 0pt, clip, width=0.35\textwidth]{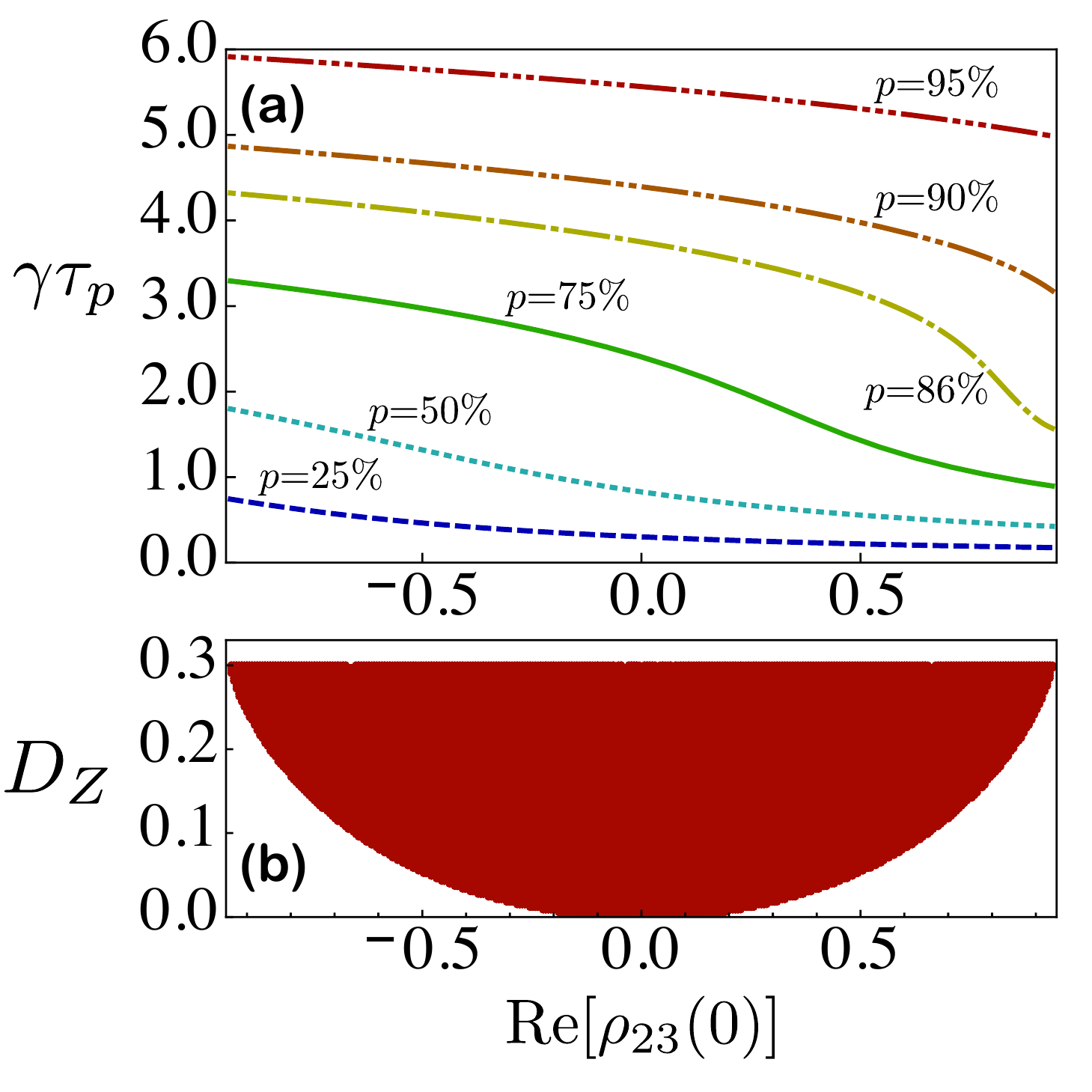}
	\caption{(Color online) (a): $\gamma \tau_p$ against ${\rm Re}[\rho_{23}(0)]$ for $p\ug95$ (red three-dotted-dashed line), $p \ug 90$ (orange two-dotted-dashed), $p\ug86$ (yellow dot-dashed), $p\ug75$ (green solid), $p\ug50$ (cyan dotted) and $p\ug25$ (blue dashed).  (b): Quantum discord as a function of ${\rm Re}[\rho_{23}(0)]$. Throughout, we have set  $\xi_{S}\ug0.25$ and $N\ug0$.}
	\label{FigTime2}
	\end{center}
\end{figure}
\subsection{Influence of initial quantum correlations} \label{Qubits-in}

In line with Subsection \ref{HO-ent}, we next investigate the connection between heat flux and typical measures of correlations of the initial state (\ref{rhoINI}). These measures, namely the concurrence for entanglement and the quantum discord for general non-classical correlations, are described in Appendix \ref{AppE}. Unlike family (\ref{CV-family}) for harmonic oscillators, all the qubit states (\ref{rhoINI}) are {\it disentangled} (as can be shown by explicitly calculating the concurrence \cite{wootters}, see Appendix \ref{AppE}). They all feature, however, some quantum discord $D_Z$. To show this, in figure 5(c) we set $\xi_{S}\ug0.25$ and plot $D_Z $ \cite{zurek2001,vedral2010} as a function of ${\rm Re}[\rho_{23}(0)]$ and ${\rm Im}[\rho_{23}(0)]$. Similarly to the behavior of $D_G$ in figure 5(a), $D_Z $ is non-zero on the entire plane but the origin ${\rm Re}[\rho_{23}(0)]\ug{\rm Im}[\rho_{23}(0)]\ug0$. In the present case, a simpler functional dependance arises since $D_Z $ depends only on $|\rho_{23}(0)|$ and it is thus constant along each circle centred at the origin. As $|\rho_{23}(0)|$ grows up, $D_Z $ increases.

We see that, similarly to harmonic oscillators, states with different discord can exhibit the same heat flux dynamics [corresponding to a set value of ${\rm Re}[\rho_{23}(0)]$]. To better highlight this, in figure 6(b) for a fixed value of of ${\rm Re}[\rho_{23}(0)]$, we report all the possible values of  $D_Z$. Similarly to the harmonic oscillators case, we see that the slowest and fastest heat flows occur only for the maximum value of discord. However, a high amount of discord does not necessarily imply a low or fast energy release as witnessed by the fact that states with maximum value of $D_Z$ are compatible with any heat flux dynamics.

The above indicates that, also in the case of qubits, it is the structure of correlations that decides the speed of heat flux.

\section{Interpretation of correlated heat flux for qubits}

The non-local nature of the correlated heat $\mathcal J_{12}$ [\cf\eqs(\ref{Jt}) and (\ref{J12})] suggests a possible link between such quantity and some measure of correlations between $S_1$ and $S_2$. A general formulation of such a connection with some known correlations indicator is not straightforward. Remarkably, however, we next find that, in the case of qubits, this is possible for a relevant class of initial states. Specifically, we show that $\mathcal J_{12}$ can be expressed in terms of the so called {\it trace distance discord} (TDD) \cite{debarba} whenever $S$ is initially in a product of local thermal states. This is a well-behaved measure of non-classical correlations exhibited by a bipartite quantum state (not necessarily in the presence of entanglement). Specifically, the one-sided  trace distance discord (TDD) $\mathcal{D}_{\rightarrow}(\rho)$ from $1$ to $2$ of a bipartite quantum
state $\rho$ is defined as the minimal {\it trace norm} distance \cite{nc} between such state and the set of so called {\it classical-quantum} (CQ) states \cite{debarba}. A CQ state features zero QCs with respect to local measurements on $A$ and can be expressed as
\begin{eqnarray}
\rho_{CQ} = \sum_j |\alpha_j\rangle_1\langle \alpha_j|\otimes \varrho_2(j)\;
 \label{CQSTATES}\end{eqnarray}  
with  $\{|\alpha_j\rangle_1\}$ being a complete set of orthonormal vectors of subsystem $1$ and $\varrho_2(j)$ being a positive (not necessarily normalized) operator of subsystem $2$.
Specifically, if $\| \Theta\|_1 = \mbox{Tr}[ \sqrt{\Theta^\dag \Theta}]$ denotes the trace norm (or Schatten 1-norm) of a generic operator $\Theta$ then the TDD of state $\rho$ is defined by
\begin{eqnarray}
\mathcal{D}_{{\rightarrow}}(\rho) =\frac{1}{2}    \min_{\{\rho_{CQ}\}}\! \| \rho-\rho_{CQ}\|_1 \;, \label{formaldef}
\end{eqnarray} 
where, as shown by the notation, the minimum is over all possible quantum-classical states \eqref{CQSTATES}. In other words, the TDD is the minimum distance in the Hilbert space between between $\rho$ and the set of CQ states.

We next restrict to a system $S$ made out of a pair of qubits and initially in the state $\rho(0){=}\frac{\exp[{-{\hat{H}_1}}/(k_{\rm B}T_1)]}{Z_1}\otimes\frac{\exp[{-{\hat{H}_2}}/(k_{\rm B}T_2)]}{Z_2}$, namely a tensor product of two local thermal states (in general at different temperatures). Note that such a family encompasses the initial state considered in Subsection \ref{thermal-qubits} as a special case. Using the solution of the master equation given in Appendix \ref{AppC}, it can be easily shown that the state of $S$ will maintain the same form at any time (the local temperatures can change with time). A locally thermal state belongs to the family of two-qubit $X$ states (these have non-zero entries on the two main diagonals of the corresponding density matrix). The TDD of such states can be calculated exactly \cite{ciccarello}. Using the closed formula of Ref.~\cite{ciccarello}, we find that at any time $t$ the modulus of the correlated heat flux $\mathcal{J}_{12}(t)$ is proportional to the TDD of state $\rho(t)$ according to
\begin{equation}
	\left|\mathcal{J}_{12}(t)\right| \ug 4\hbar \omega \gamma\xi\, \mathcal D_{\rightarrow}[\rho(t)]	\,.\label{EqNonLocalFluxDiscord}
\end{equation}
It is natural to wonder whether this property holds for more general initial states. This is not the case as can be seen through the following counterexample: let us select the initial state
\begin{align}
	\rho(0) =\frac{1}{2} \left| \psi \right>_1 \left< \psi \right| \otimes \begin{pmatrix} 1-\xi_2 & 0 \\0 & 1+\xi_2 \end{pmatrix}
	\label{EqInitialRhoQubits2}
\end{align}
with 
\begin{align}
	\left| \psi \right>_1 = \sqrt{\frac{1+\xi}{2}} \left| 0 \right>_1 + \sqrt{\frac{1-\xi}{2}} \left| 1 \right>_1\,,
	\label{EqChoice}
\end{align}
where $\xi_2$ is the same as in \eq(\ref{xi}) for $T\ug T_2$. As in the previous case, $\rho(0)$ is a product state, hence featuring zero correlations, with $S_2$ locally in a thermal state. Now, however, despite having the same populations and energy as the thermal state corresponding to $\xi$, the initial state of $S_1$ is fully {\it pure}. In other words, $S_1$ has the same temperature as $R$ but features non-zero coherences.
In such a case, we can show that $\mathcal D_{\rightarrow}(t)$ is in general finite but $\mathcal{J}_{1}(t)\ug\mathcal{J}_{12}(t)\ug0$ identically. In other words, the interaction mediated by the reservoir gives rise to QCs between the system's subparts with no simultaneous development of any correlated heat flux. 
%
%

\section{Conclusions}\label{Conclusions}

In this work, we have studied the dynamics of heat flux of a bipartite system interacting with a thermal reservoir in a cascaded way. 
The cascading makes one of the two subsystems interact with the reservoir modified by the previous interaction with the other subsystem. Because of such circumstance, the local dynamics of the second subsystem is non-Markovian despite the fact that the joint dynamics is Markovian. This affects the heat flux in such a way that it exhibits a non-exponential time behaviour. We have carried out a systematic analysis of this after showing that the total heat flux can be decomposed into three components. In particular, one of these -- arising from a non-local dissipator entering the master equation -- can be identified as a correlated heat flux and was shown to play a major role in the non-monotonic time evolution. 

Typical behaviours, occurring in the case of both thermal and correlated initial states, have been scrutinized for two paradigmatic systems: a pair of harmonic oscillators with a reservoir of bosonic modes and two qubits with a reservoir of fermionic modes. While in the case of harmonic oscillators basically all of the observed features can be explained analytically, an analogous analysis is not possible for qubits. Notwithstanding, most of the qualitative features of the heat flux dynamics are quite similar to those occurring for harmonic oscillators (aside from saturation effects owing to the presence of only two levels for qubits).

In the case of thermal initial states, we have shown that the total heat flux exhibits a monotonic, although non-exponential, time behaviour. In particular, an almost flat profile arises at intermediate times which is mostly due 
to the occurrence of the aforementioned correlated heat flux. To explore the effect of initial correlations in the system state, we have focused on a suitable family of initial states that are locally thermal but 
additionally feature non-local correlations. In general, the effect of these is to cause non-monotonicity of the total heat flux accompanied by a simultaneous slow down or speed up of the thermalisation process. We have investigated the role played by the initial amount of quantum correlations, either in the form of entanglement or discord, on the rate of energy exchange. Our analysis indicates that, although the states featuring the slowest and fastest heat flux dynamics are characterized by high values of discord, it is mostly the peculiar {\it structure} of initial correlations that matters rather than their overall amount.

Finally, we have found that -- in the case of qubits and for initial thermal states -- the magnitude of the correlated heat flux at any time coincides (up to a proportionality factor) with the trace distance discord of the open system. In particular, this shows the existence of a physical scenario within which such a {\it bona fide} measure of quantum correlations acquires a clear physical significance.

It is worth to emphasise that, as already observed, a key feature of our system is that while the joint dynamics of $S_1$ and $S_2$ is Markovian, the reduced dynamics of system $S_2$ is non-Markovian. Recently, the concept of quantum non-Markovianity has received remarkable attention \cite{NM} in the effort of defining on a rigorous basis the distinctive aspects of such phenomenon and, accordingly, ways to quantify it \cite{NM2}. Within this framework, our work suggests an interesting connection between quantum non-Markovianity and heat flux dynamics.

In this work, we have focused on initial states -- either correlated or not  -- that are in any case locally thermal at a {\it uniform} temperature (i.e., the same for both subsystems). Allowing for a non-uniform temperature makes the heat flux dynamics as well as its interplay with initial correlations considerably richer, which will be the subject of a future work \cite{farace}.

\section*{Aknowledgements}
The authors are grateful to R. Fazio for useful discussions. 
This work is funded by the EU Collaborative Project TherMiQ (Grant Agreement 618074) and the Italian PRIN-MIUR 2010/2011.

\begin{appendix}

\section{Stationary state}
\label{AppA}

Here, we prove that the thermal state (\ref{rho-infty}) is indeed the asymptotic state reached by $S$ both in the case of harmonic oscillators and qubits. Let $\rho_{\rm th}\ug e^{- \beta\hat H_1}e^{- \beta\hat H_2}$ with $\beta\ug 1/(k_{\rm B}T)$ (the tensor product symbol is omitted for simplicity). To demonstrate that this is indeed the system's steady state, we will prove that $\rho_{\rm th}$ fulfils the master equation under stationary conditions (when all the time derivatives vanish), namely 
\begin{eqnarray}
(\mathcal L_1\piu\mathcal L_2\piu\mathcal D_{12})(\rho_{\rm th})\ug0\,.
\end{eqnarray}
\subsubsection{Harmonic oscillators}
Let $\hat{\mathcal {U}}_\pm\ug e^{\pm\beta \hbar \omega \hat a^\dagger \hat a}$, where $\hat a$ and $\hat a^\dag$ are bosonic annihilation and creation operators. Then, $\hat{\mathcal {U}}_- \hat a \,\hat{\mathcal {U}}_+\ug e^{\beta \hbar \omega} \; \hat a$ and $\hat{\mathcal {U}}_- \hat a^\dag \,\hat{\mathcal {U}}_+\ug e^{-\beta \hbar \omega} \; \hat a^\dag$. These identities entail 
\begin{eqnarray}
 \left[e^{-\beta \hbar \omega \hat a^\dagger \hat a}, \hat a \right] = \left(1 \meno e^{\meno \beta \hbar \omega} \right) e^{\meno \beta \hbar \omega \hat a^\dagger \hat a} \; \hat a,\,\label{Id1}\\
  \left[e^{\meno \beta \hbar \omega \hat a^\dagger \hat a}, \hat a^\dag \right] = \left(1 \meno e^{\beta \hbar \omega} \right) e^{\meno \beta \hbar \omega \hat a^\dagger \hat a} \; \hat a^\dag\,.\label{Id2}
\end{eqnarray}
In the present case, $\rho_{\rm th}\ug e^{\meno \beta \hbar \omega \hat a_1^\dagger \hat a_1}e^{\meno \beta \hbar \omega \hat a_2^\dagger \hat a_2}$.  Applying $\mathcal L_1$ [\cf\eq\eqref{Li1}] to such a state, upon use of \eqs(\ref{Id1}) and (\ref{Id2}), yields
\begin{widetext}
\begin{eqnarray}
\mathcal L_1(\rho_{\rm th})&\ug&\left[\gamma (N\piu 1) \Big( e^{\meno \beta \hbar \omega \hat a_1^\dagger \hat a_1} e^{\meno \beta \hbar \omega} \hat a_1 \hat a_1^\dagger \meno  e^{\meno \beta \hbar \omega \hat a_1^\dagger \hat a_1} \hat a_1^\dagger \hat a_1 \Big)
\piu  \gamma N\Big( e^{\meno \beta \hbar \omega \hat a_1^\dagger \hat a_1} e^{\beta \hbar \omega}\hat a_1^\dagger  \hat a_1 \meno  e^{\meno \beta \hbar \omega \hat a_1^\dagger \hat a_1} \hat a_1^\dagger \hat a_1 \meno  e^{\meno \beta \hbar \omega \hat a_1^\dagger \hat a_1} \Big) \nonumber\right] e^{\meno \beta \hbar \omega \hat a_2^\dagger \hat a_2}\nonumber \\
&\ug& \left[\gamma (N\piu 1) (e^{\meno \beta \hbar \omega} \meno 1 ) \hat a_1^\dagger \hat a_1 \piu \gamma (N\piu 1) e^{\meno \beta \hbar \omega} \piu  \gamma N   (e^{\beta \hbar \omega}\meno 1) \hat a_1^\dagger \hat a_1 \meno  \gamma N   \right]\rho_{\rm th} \nonumber
\\\nonumber\\
&\ug&(\meno  \gamma\hat a_1^\dagger \hat a_1 \piu \gamma N\piu  \gamma \hat a_1^\dagger \hat a_1 \meno  \gamma N) \rho_{\rm th} =0\,.\nonumber
\end{eqnarray}
\end{widetext}
Likewise, the identity $\mathcal L_2(\rho_{\rm th})\ug0$ is proven by swapping indexes 1 and 2. The last step is thus showing that $\mathcal D_{12}\rho_{\rm th}\ug0$ (\cf\eq(\ref{D121})). Using again eqs (\ref{Id1}) and (\ref{Id2}) gives
\begin{widetext}
\begin{eqnarray}
\mathcal D_{12}(\rho_{\rm th})&\ug&\left\{ \gamma (N\piu 1) \Big[ e^{\meno \beta \hbar \omega} (1\meno e^{\beta \hbar \omega})  \hat a_1 \hat a_2^\dagger \meno   (1\meno e^{\meno \beta \hbar \omega}) \hat a_1^\dagger \hat a_2 \Big]\piu \gamma N \Big[ e^{\beta \hbar \omega} (1\meno e^{\meno \beta \hbar \omega}) \hat a_1^\dagger \hat a_2 \meno   (1\meno e^{\beta \hbar \omega}) \hat a_1\hat a_2^\dagger \Big]\right\}\rho_{\rm th}\nonumber\\
&\ug& \Big[ \gamma (N\piu 1) e^{\meno \beta \hbar \omega} (1\meno e^{\beta \hbar \omega}) \hat a_1 \hat a_2^\dagger \meno  \gamma N  (1\meno e^{\beta \hbar \omega})\hat a_1\hat a_2^\dagger \meno    \gamma (N\piu 1) (1\meno e^{\meno \beta \hbar \omega}) \hat a_1^\dagger \hat a_2 \piu \gamma N e^{\beta \hbar \omega} (1\meno e^{\meno \beta \hbar \omega})  \hat a_1^\dagger \hat a_2  \Big] \rho_{\rm th}\nonumber\\
&\ug&  \Big[ \gamma N(1\meno e^{\beta \hbar \omega}) \hat a_1 \hat a_2^\dagger \meno  \gamma N  (1\meno e^{\beta \hbar \omega})\hat a_1\hat a_2^\dagger\meno    \gamma (N\piu 1) (1\meno e^{\meno \beta \hbar \omega}) \hat a_1^\dagger \hat a_2 \piu \gamma (N\piu 1) (1\meno e^{\meno \beta \hbar \omega})  \hat a_1^\dagger \hat a_2  \Big] \rho_{\rm th}\ug 0\,.\nonumber
\end{eqnarray}
\end{widetext}
This concludes the proof.
\subsubsection{Qubits}
In this case $\rho_{\rm th}\ug e^{-\beta \hat H_{1}} e^{-\beta \hat H_{2}}/Z^2$, which we rearrange as $\rho_{\rm th}\ug\rho_{1{\rm th}}\rho_{2{\rm th}}$ with
\begin{equation}
\rho_{i\rm th}{=}\frac{1}{Z} \left(\frac{\openone_i}{2}-\frac{\xi}{2}\hat\sigma_{iz}\right)\,.
\end{equation}
Using $\hat\sigma_{j\pm}\hat\sigma_{jz}\sigma_{j\mp}\ug{\mp}\sigma_{j\pm}\hat\sigma_{j\mp}$ and $\hat\sigma_{jz}\hat\sigma_{j\pm}\hat\sigma_{j\mp}\ug\hat\sigma_{j\pm}\hat\sigma_{j\mp}\sigma_{jz}\ug{\pm}\hat\sigma_{j\pm}\hat\sigma_{j\mp}$ it is immediate to see that $\mathcal{L}_i(\rho_{\rm th})\ug0$ [\cf\eqref{Li2}] since $\mathcal{L}_i(\openone_i)\ug\xi\mathcal{L}_i(\hat\sigma_{iz})$.\\
On the other hand, from \eq\eqref{D122} follows
\begin{eqnarray}
\mathcal{D}_{12}(\rho_{\rm th})&\ug&\hat\sigma_{1-}
\frac{\gamma}{2}\left[\hat\sigma_{1-},\rho_1^{\rm th}\right] \left[\rho_2^{\rm th},\hat\sigma_{2+}\right]\!+\!\frac{\gamma}{2}\left[\hat\sigma_{1+},\rho_1^{\rm th}\right] \left[\rho_2^{\rm th},\hat\sigma_{2-}\right]\nonumber\\
&&+\frac{\gamma\xi}{2} \lbrace\hat\sigma_{1-} ,\rho_1^{\rm th}\rbrace\left[\rho_2^{\rm th},\hat\sigma_{2+}\right]\!-\!\frac{\gamma\xi}{2} \lbrace\hat\sigma_{1+} ,\rho_1^{\rm th}\rbrace\left[\rho_2^{\rm th},\hat\sigma_{2-}\right]\!,\nonumber
\end{eqnarray}
which upon use of $[\hat\sigma_k^\pm,\rho_k^{\rm th}]{\!=}{\pm}\xi \hat\sigma_k^{\pm}$ and $\lbrace\hat\sigma_k^\pm,\rho_k^{\rm th}\rbrace{\ug}\hat\sigma_k^{\pm}$ yields
\begin{eqnarray}
\mathcal{D}^{12}[\rho_{\rm th}]=
&\frac{\gamma\xi}{2}\left(-\hat\sigma_1^{-}\right) \left(-\xi\hat\sigma_2^{+}\right){\!+\!}\frac{\gamma\xi}{2}\left(\hat\sigma_1^{+}\right) \left(\xi\hat\sigma_2^{-}\right)\nonumber\\
&{+}\frac{\gamma}{2}\xi \left(\hat\sigma_1^{-} \right)\left(-\xi \hat\sigma_2^{+}\right){\!-\!}\frac{\gamma\xi}{2}\left(\hat\sigma_1^{+} \right)\left(\xi \hat\sigma_2^{-}\right)\ug0\,.\,\,
\end{eqnarray}
This concludes the proof.

\section{Time evolution of the covariance matrix for harmonic oscillators}\label{AppB}

For a given initial state, the explicit calculation of the coefficients $C_{mn}(t)$ entering the heat fluxes in \eqs\eqref{J1tcv}-\eqref{J12tcv} is conveniently carried out through the Langevin equations \cite{Gardiner1994a}.
These are equivalent to the master equation \eqref{ME} and read
\begin{align}
	\frac{d}{dt} \! \begin{pmatrix} \hat X_1 \\ \hat Y_1 \\ \hat X_2 \\ \hat Y_2 \end{pmatrix} \!=\! -\gamma\begin{pmatrix} \tfrac{1}{2} & 0 &0 & 0 \\ 0 & \tfrac{1}{2} & 0 & 0 \\ 1 & 0 & \tfrac{1}{2} & 0 \\ 0 & 1 & 0 & \tfrac{1}{2} \end{pmatrix} \! \begin{pmatrix} \hat X_1 \\ \hat Y_1 \\ \hat X_2 \\ \hat Y_2 \end{pmatrix} \! - \sqrt{\gamma} \begin{pmatrix} \hat X_{\rm in} \\ \hat Y_{\rm in} \\ \hat X_{\rm in} \\ \hat Y_{\rm in} \end{pmatrix},
	\label{EqLangevinOscillator}
\end{align}\\
\normalsize
where $\hat X_{\rm in}$ and $\hat Y_{\rm in}$ are zero-mean Gaussian noises characterized by the correlations $\left< \hat X_{\rm in} \hat Y_{\rm in} \right> {=} 0$, $\left< \hat X_{\rm in} \hat X_{\rm in} \right>{=}\left< \hat Y_{\rm in} \hat Y_{\rm in} \right> {=} N{ +} \tfrac{1}{2}$. 
Correspondingly, the covariance matrix evolves in time as
\begin{eqnarray}
\quad \frac{d}{dt} C = A C + C A^T + M,
	\label{EqMomentsXY}
\end{eqnarray}
where $A$ is the matrix appearing in Eq.~\eqref{EqLangevinOscillator} and
\begin{align}
	M = \gamma \left(N + \tfrac{1}{2}\right) \begin{pmatrix} 1 & 0 & 1 & 0\\ 0 &1 & 0 & 1 \\ 1 & 0 & 1 & 0\\ 0 & 1 & 0 & 1 \end{pmatrix}.
	\label{EqThermalM}
\end{align}
The solution of such a linear first-order differential system yields the covariance matrix vs.~time and, in particular, the time-dependent coefficients appearing in \eqs\eqref{J1tcv}-\eqref{J12tcv}. The relevant equations are\\
\begin{align}
\dot{C}_{11}(t)&=-\gamma \left[ C_{11}(t) - (N+\tfrac{1}{2})\right], \label{C11}\\
\dot{C}_{22}(t)&=-\gamma \left[ C_{22}(t) - (N+\tfrac{1}{2})\right], \label{C22}\\
\dot{C}_{33}(t)&=-\gamma \left[ C_{33}(t) - (N+\tfrac{1}{2})\right] - 2 \gamma C_{13}(t), \label{C33}\\
\dot{C}_{44}(t)&=-\gamma \left[ C_{44}(t) - (N+\tfrac{1}{2})\right] - 2 \gamma C_{24}(t), \label{C44}\\
\dot{C}_{13}(t)&=-\gamma C_{13}(t) - \gamma \left[ C_{11}(t) - (N+\tfrac{1}{2})\right], \label{C13}\\
\dot{C}_{24}(t)&=-\gamma C_{24}(t) - \gamma \left[ C_{22}(t) - (N+\tfrac{1}{2})\right], \label{C24}
\end{align}
\begin{align}
\dot{C}_{12}(t)&=-\gamma C_{12}(t), \label{C12}\\
\dot{C}_{14}(t)&=-\gamma C_{14}(t) - \gamma C_{12}(t), \label{C14}\\
\dot{C}_{23}(t)&=-\gamma C_{23}(t) - \gamma C_{12}(t), \label{C23}\\
\dot{C}_{34}(t)&=-\gamma C_{34}(t) -\gamma C_{14}(t) -\gamma C_{23}(t).  \label{C34}
\end{align}
We thus find two independent families of equations: one for the $\left< X_i X_j\right>$, $\left< Y_i Y_j\right>$ correlations and one for $\left< X_i Y_j\right>$. In particular, \eqs\eqref{C11}-\eqref{C24} completely determine the evolution of the heat flux as can be seen upon inspection of \eqs\eqref{J1tcv}-\eqref{J12tcv}.

\section{Time evolution of the density matrix for qubits}\label{AppC}

In the Liouville space \cite{mukamel}, the density operator of the two qubits $S_1$ and $S_2$ reads
\begin{equation}
\rho(t)=\sum_{kj}{\rm Tr}[\rho(t)\ket{j}\bra{k}]\ket{k}\bra{j} = \sum_{kj}\rho_{kj}(t)\kket{kj}
\end{equation}
with $k,j\ug1,...,4$, $|1\rangle\!\equiv\!|ee\rangle_{12}$, $|2\rangle\!\equiv\!|eg\rangle_{12}$ , $|3\rangle\!\equiv\!|ge\rangle_{12}$  and $|4\rangle\!\equiv\!|gg\rangle_{12}$ and where we have adopted a double-bracket notation according to which $\kket{kj}\!\equiv\!\ket{k}\!\!\bra{j}$ is a vector in the Liouville space vector. Hence, in such a space $\rho$ is a vector expressed as a linear combination of the basis vectors $\{\kket{kj}\}$ (vectorization). Accordingly, master equation (\ref{ME}) can be written in the matrix form $\dot{\rho}=\mathcal{K}\rho$, where matrix $\mathcal K$ is defined by $\mathcal{K}_{kj,mn}\ug\bbra{kj}\mathcal{L}\kket{mn}\ug{\rm Tr}\lbrace\ket{j}\bra{k}\mathcal{L}(\ket{m}\bra{n})\rbrace$. In our case, such matrix is explicitly given by 
\begin{widetext}
\begin{equation}
\frac{\mathcal{K}}{2\gamma}\ug\left(
\begin{smallmatrix}
 {-}2 (1\piu\xi)& 0 & 0 & 0 & 0 & 1{-}\xi  & 1{-}\xi  & 0 & 0 & 1{-}\xi  & 1{-}\xi  & 0 & 0 & 0 & 0 & 0 \\
 0 & -2\meno\xi & \xi {-}1 & 0 & 0 & 0 & 0 & 1{-}\xi  & 0 & 0 & 0 & 1{-}\xi  & 0 & 0 & 0 & 0 \\
 0 & -1\meno\xi & -2\meno\xi & 0 & 0 & 0 & 0 & 1{-}\xi  & 0 & 0 & 0 & 1{-}\xi  & 0 & 0 & 0 & 0 \\
 0 & 0 & 0 & {-}2 & 0 & 0 & 0 & 0 & 0 & 0 & 0 & 0 & 0 & 0 & 0 & 0 \\
 0 & 0 & 0 & 0 & -2\meno\xi & 0 & 0 & 0 & \xi {-}1 & 0 & 0 & 0 & 0 & 1{-}\xi  & 1{-}\xi  & 0 \\
 1\piu\xi & 0 & 0 & 0 & 0 & {-}2 & \xi {-}1 & 0 & 0 & \xi {-}1 & 0 & 0 & 0 & 0 & 0 & 1{-}\xi  \\
 1\piu\xi & 0 & 0 & 0 & 0 & -1\meno\xi & {-}2 & 0 & 0 & 0 & \xi {-}1 & 0 & 0 & 0 & 0 & 1{-}\xi  \\
 0 & 1\piu\xi & 1\piu\xi & 0 & 0 & 0 & 0 &  -2\piu\xi & 0 & 0 & 0 & \xi {-}1 & 0 & 0 & 0 & 0 \\
 0 & 0 & 0 & 0 & -1\meno\xi & 0 & 0 & 0 & -2\meno\xi & 0 & 0 & 0 & 0 & 1{-}\xi  & 1{-}\xi  & 0 \\
 1\piu\xi & 0 & 0 & 0 & 0 & -1\meno\xi & 0 & 0 & 0 & {-}2 & \xi {-}1 & 0 & 0 & 0 & 0 & 1{-}\xi  \\
 1\piu\xi & 0 & 0 & 0 & 0 & 0 & -1\meno\xi & 0 & 0 & -1\meno\xi & {-}2 & 0 & 0 & 0 & 0 & 1{-}\xi  \\
 0 & 1\piu\xi & 1\piu\xi & 0 & 0 & 0 & 0 & -1\meno\xi & 0 & 0 & 0 &  -2\piu\xi & 0 & 0 & 0 & 0 \\
 0 & 0 & 0 & 0 & 0 & 0 & 0 & 0 & 0 & 0 & 0 & 0 & {-}2 & 0 & 0 & 0 \\
 0 & 0 & 0 & 0 & 1\piu\xi & 0 & 0 & 0 & 1\piu\xi & 0 & 0 & 0 & 0 &  -2\piu\xi & \xi {-}1 & 0 \\
 0 & 0 & 0 & 0 & 1\piu\xi & 0 & 0 & 0 & 1\piu\xi & 0 & 0 & 0 & 0 & -1\meno\xi &  -2\piu\xi & 0 \\
 0 & 0 & 0 & 0 & 0 & 1\piu\xi & 1\piu\xi & 0 & 0 & 1\piu\xi & 1\piu\xi & 0 & 0 & 0 & 0 & 2 (\xi {-}1) \nonumber
\end{smallmatrix}
\right)\,,
\end{equation}
\end{widetext}
where we have used the ordering
\begin{equation}
\left(
\begin{array}{cccccccc}
 \mathcal{K}_{11,11} & \mathcal{K}_{11,12} & \mathcal{K}_{11,13} & \mathcal{K}_{11,14} & \mathcal{K}_{11,21} & \cdots\\
 \mathcal{K}_{21,11} & \mathcal{K}_{21,12} & \mathcal{K}_{21,13} & \mathcal{K}_{21,14} & \mathcal{K}_{21,21} & \\
 \mathcal{K}_{31,11} & \mathcal{K}_{31,12} & \mathcal{K}_{31,13} & \mathcal{K}_{31,14} & \mathcal{K}_{31,21} & \\
 \mathcal{K}_{41,11} & \mathcal{K}_{41,12} & \mathcal{K}_{41,13} & \mathcal{K}_{41,14} & \mathcal{K}_{41,21} &  \\
 \mathcal{K}_{12,11} & \mathcal{K}_{12,12} & \mathcal{K}_{12,13} & \mathcal{K}_{12,14} & \mathcal{K}_{12,21} & \\
 \vdots & & & & &\ddots \,.
\end{array}
\right)\,.
\end{equation}
The solution of the linear first-order differential system $\dot{\rho}=\mathcal{K}\rho$ is found in an exponential form as 
\begin{equation}
\rho_{mn}(t)=\sum_{k,j}\left(e^{\mathcal{K}t}\right)_{mn,kj}\rho_{kj}(0)\,.
\end{equation}
In particular, it turns out that
\begin{equation}
\rho_{14}(t)=e^{-\gamma t}\rho_{14}(0)\,,
\end{equation}
which shows that the off-diagonal terms $\rho_{14}(t)\ug\rho_{41}(t)^*$ are decoupled from other elements of the density matrix regardless of the system initial state.

\section{Parametrization of initial correlated states for harmonic oscillators}
\label{AppD}

As discussed in the main text (Section \ref{SubsecCorrHeat}), in the case of harmonic oscillators we focus on the family of initial states whose associated covariance matrix reads
\begin{align}
	C(0)=
	\begin{pmatrix}
		C_{11}(0) & 0 & C_{13}(0) & 0 \\
		0 & C_{11}(0) & 0 & C_{24}(0) \\
		C_{13}(0) & 0 & C_{11}(0) & 0 \\
		0 & C_{24}(0) & 0 & C_{11}(0)
	\end{pmatrix},
	\label{EqInitialOscillatorRequests}
\end{align}
where $C_{11}(0) = N_S + \tfrac{1}{2}$ and the total energy is fixed to $U\ug\tfrac{1}{2} {\rm Tr}[C(0)]=2 C_{11}(0)$. This choice is motivated by the fact that the heat flux depends only on $C_{ii}(t)$, $C_{13}(t)$, $C_{24}(t)$ (see Eqs. \eqref{J1tcv}-\eqref{J12tcv}) and these instantaneous values are completely determined by the initial conditions $C_{ii}(0)$, $C_{13}(0)$, $C_{24}(0)$ (see Eqs. \eqref{C11}-\eqref{C24}). We could then choose any value for the remaining off-diagonal terms without affecting the heat flux, but the optimal choice is zero, as argued at the end of the section. Our essential task is to derive the conditions on the off-diagonal elements $C_{13}(0)$ and $C_{24}(0)$, in order for $C(0)$ to describe a physical state once the total energy is fixed. In general, a covariance matrix of a physically admissible Gaussian state must be such that all the second moments fulfill the Heisenberg uncertainty relations. This requirement can be expressed compactly as the semi-positivity condition
\begin{align}
	C(0) + \frac{i}{2} \begin{pmatrix} 0 & 1 & 0 & 0 \\ -1 & 0 & 0 & 0 \\ 0 & 0 & 0 & 1 \\ 0 & 0 & -1 & 0 \end{pmatrix} \geq 0.
	\label{EqHeis}
\end{align}
This is equivalent to two necessary and sufficient conditions~\cite{Pirandola2009a}. First, the covariance matrix needs to be positive, i.e., $C(0)>0$, which is in turn equivalent to the two inequalities
\small
\begin{align}
	\left| C_{13}(0) \right|<C_{11}(0) = N_S + \tfrac{1}{2}, \quad \left| C_{24}(0) \right|<N_S + \tfrac{1}{2}.
	\label{EqPositivityCov}
\end{align}
\normalsize
Second, the symplectic eigenvalues $\nu_{\pm}$ must fulfil
\footnotesize
\begin{align}
	\nu_{\pm} = \sqrt{\frac{I_A + I_B + 2I_C \pm \sqrt{(I_A + I_B + 2I_C)^2-4I_\Sigma}}{2}} \geq \frac{1}{2},
	\label{EqSymp}
\end{align}
\normalsize
where we introduced the symplectic invariants~\cite{Ferraro2005a} $I_A\ug I_B\ug C_{11}(0)^2$, $I_C \ug C_{13}(0) C_{24}(0)$ and $I_\Sigma \ug C_{11}(0)^4 \piu C_{13}(0)^2C_{24}(0)^2 \meno C_{11}(0)^2[C_{13}(0)^2 \piu C_{24}(0)^2]$. Note that if the pair $\{C_{13}(0),C_{24}(0)\}\ug \{c,d\}$ satisfies the two conditions, so do the pairs $\{C_{13}(0),C_{24}(0)\}\ug\{d,c\}$ and $\{C_{13}(0),C_{24}(0)\}\ug\{-c,-d\}$. Hence, the region of physicality is symmetric across the two diagonals of the $C_{13}(0)-C_{24}(0)$ plane. In figure 7, we plot this region for different values of $N_S$.
\begin{figure}[htbp]
	\begin{center}
	\includegraphics[trim=0pt 0pt 0pt 0pt, clip, width=0.37\textwidth]{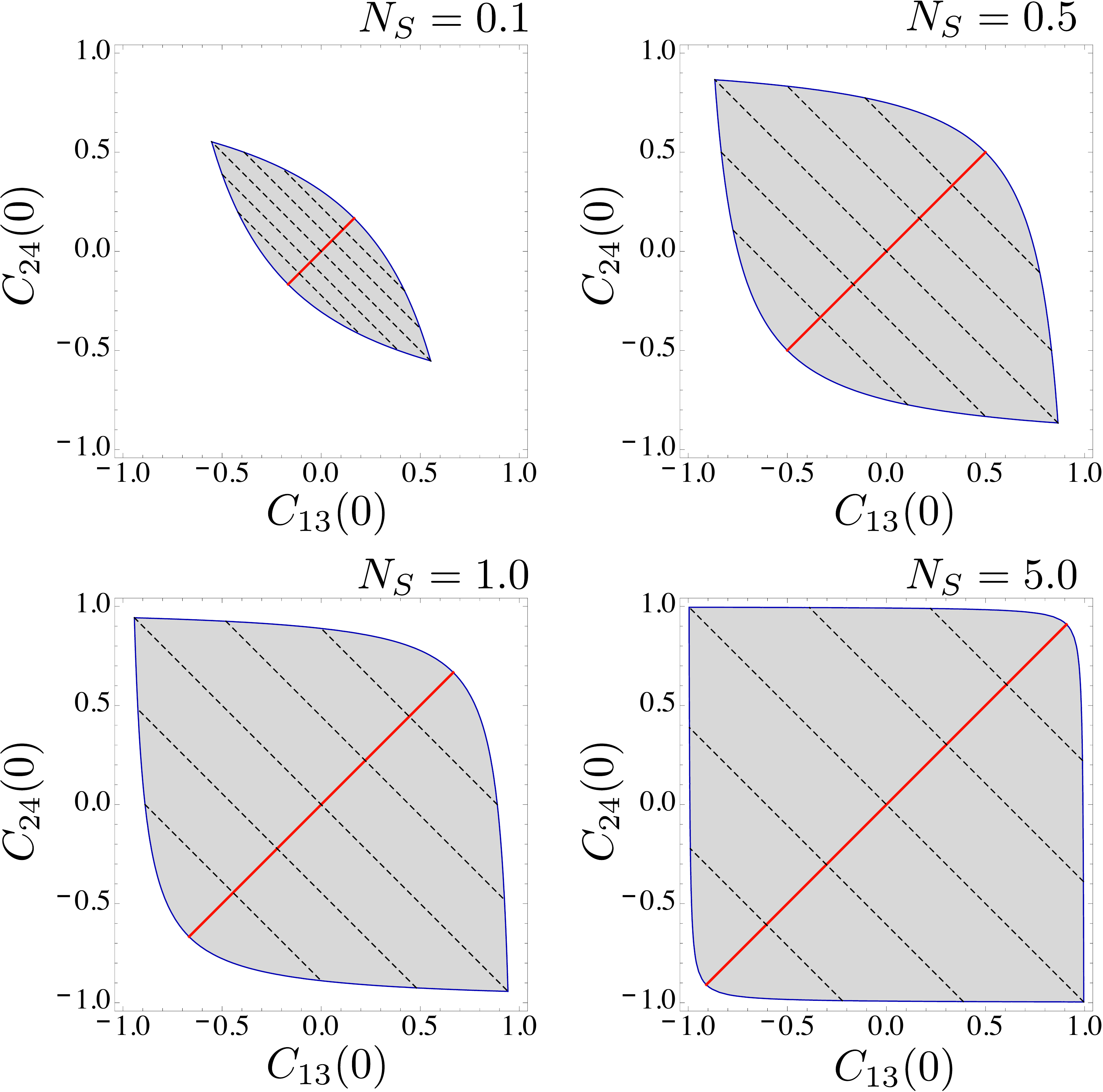}
	\caption{(Color online) Domain on the $C_{13}(0)-C_{24}(0)$ plane within which the covariance matrix represents a physical Gaussian state for different values of $N_S$, i.e., the total energy. Both $C_{13}(0)$ and  $C_{24}(0)$ are expressed in units of $N_S\piu 1/2$. The sum $C_{13}(0)\piu C_{24}(0)$ is constant along each black dashed line. The red dashed line is instead the set of points such that $C_{13}(0)\ug C_{24}(0)$.}
	\label{FigValidity}
	\end{center}
\end{figure}
One can see that the area of the physicality region grows with $N_S$. Indeed, if $N_S\ug 0$ each local state [i.e., $\rho_1(0)$ and $\rho_2(0)$] is pure since $S$ is in the vacuum state, hence no correlations are present. Moreover, note that the line where $C_{13}(0)\ug C_{24}(0)$ (red line in figure~...) spans all the allowed values of $C_{13}(0)\piu C_{24}(0)$ (this is constant along each black dashed line in the plots).
As heat fluxes depend on $C_{13}(0)$ and $C_{24}(0)$ through their sum $C_{13}(0)\piu C_{24}(0)$ [\cf\eqs(\ref{J2-corr})-(\ref{J22-corr})], we see that, in order to explore all the possible heat flux dynamics, one can set $C_{13}(0)\ug C_{24}(0)$ without loss of generality. In other words, given a black dashed line (see figure~\ref{FigValidity}), any covariance matrix lying on it  yields the same heat flux dynamics as that associated with its intersection point with the red line. 
Moreover, for states such that $C_{13}(0)\ug C_{24}(0)$ the constraints~\eqref{EqPositivityCov} and~\eqref{EqSymp} can be combined into the single condition $\left| C_{13}(0) \right|\leq N_S$. This entails that, in the light of the above considerations,
\begin{equation}
	\left| C_{13}(0) +C_{24}(0) \right|\leq 2N_S\,.
	\label{bound-sum}
\end{equation}
If we had other non-zero off-diagonal terms, the constraints \eqref{EqPositivityCov} and~\eqref{EqSymp} would be more restrictive on $C_{13}(0)$ and $C_{24}(0)$. In other words we would get $\left| C_{13}(0) +C_{24}(0) \right|\leq C_{MAX} <2N_S$ and some possible evolutions of the heat flux would remain unexplored. Starting with a state of the form \eqref{EqInitialOscillatorRequests} allows instead for a complete analysis of the problem.
\section{Role of initial correlations}\label{AppX}
Introducing the collective basis $\lbrace |ee\rangle,|\Psi^+\rangle,|\Psi^-\rangle,|gg\rangle \rbrace$, where $|\Psi^\pm\rangle\!\equiv\!1/\sqrt{2}(|eg\rangle_{12}\pm|ge\rangle_{12})$, the initial state \eqref{rhoINI} becomes
\begin{equation}
\rho(0)\ug\frac{1}{4}\!\left(
\begin{smallmatrix}
(1-\xi_{S})^2& 0 & 0 & 0\\
0  & 1\meno\xi_{S}^2\piu{\rm Re}[\rho_{23}(0)] &  0 &  0 \\
0  &  0  & 1\meno\xi_{S}^2-{\rm Re}[\rho_{23}(0)] &  0 \\
0  &  0  & 0 & (1\piu\xi_{S})^2
\end{smallmatrix}
\right).
\label{rhoINIpm}
\end{equation}
Clearly, a positive (negative) value of ${\rm Re}[\rho_{23}]$ means a smaller (larger) initial population of $|\Psi^-\rangle$ compared to the case where ${\rm Re}[\rho_{23}]\ug0$. On other hand, master equation \eqref{ME} can be reexpressed as \cite{zoller}
\begin{equation}
\dot{\rho}=-i \left[\tilde{H},\rho\right]+\tilde{\mathcal{L}}(\rho),
\end{equation}
where we have defined 
\begin{eqnarray}
\tilde{\mathcal{L}}(\rho){=}
&&\Gamma^+\mathcal{L}\big[|\Psi^+\rangle\langle gg|+|ee\rangle\langle\Psi^+|\big](\rho)+\nonumber\\
&&\Gamma^-\mathcal{L}\big[|\Psi^-\rangle\langle gg|-|ee\rangle\langle\Psi^-|\big](\rho)\,,\\
\tilde{H}=&&H-i\frac{\gamma}{2}\left(|\Psi^+\rangle\langle \Psi^-|-|\Psi^-\rangle\langle \Psi^+|\right)
\label{spincascade}
\end{eqnarray}
with $\mathcal{L}[\hat{o}](\rho){=} \hat{o}\rho\hat{o}^\dag {-}\frac{1}{2}\lbrace \hat{o}\hat{o}^\dag,\rho\rbrace$ for a generic operator $\hat{o}$ and $\Gamma^\pm=(1\pm\xi)/2$. It is clear that for $N{=}0$, i.e., $\xi{=}1$, as in the plot in fig. \ref{FigTime2}, $|\Psi^-\rangle$ is not directly affected by dissipation, which yields a slow down of energy releasing if ${\rm Re}[\rho_{23}] \!<\! 0$.
\section{Computation of quantum correlations}\label{AppE}

\subsection{Discord-like measures}

Given a pair of quantum systems $A$ and $B$, quantum discord \cite{zurek2001} is the gap between two classically equivalent expressions of the mutual information
content given by
\begin{equation}
\mathcal{D}(B|A){=}\mathcal{I}(AB){-}\mathcal{C}(B|A)\,, 
\end{equation} 
where 
\begin{equation} \label{equazionenuova}
\mathcal{I}(AB) = S(\rho_A) + S(\rho_B) - S(\rho_{AB}),
\end{equation}
is quantum mutual information \cite{popescu}, while
\begin{equation}
\mathcal{C}(B|A) = \max_{\{E_a\}} \left[ S(\rho_{B}) - \sum_a p_a S \left( \frac{{\rm Tr}_A[\rho_{AB} E_a]}{p_a} \right) \right].\label{c-cor}
\end{equation}
is interpreted as the total amount of classical correlationsIn the above expression. Here, $S(\rho)$ is the Von Neumann entropy, $\sum_a E_a \ug \openone$ is a positive-operator valued measure (POVM) on $A$ and $p_a \ug {\rm Tr}[\rho_{AB} E_a]$ is the probability of outcome $a$.

\subsubsection{Gaussian discord for harmonic oscillators}

Originally proposed for qubits, the above definition of quantum discord has been generalized to Gaussian states for continuous-variable systems \cite{Adesso2010a, Giorda2010a} under the name of Gaussian discord ${D}_G$. This is obtained by restricting the
optimization in Eq.~(\ref{equazionenuova})  
to Gaussian POVM. As a consequence ${D}_G$
provides in general only a lower bound for ${\cal D}$ 
(namely, states with non zero values of ${D}_G$ will certainly exhibit a certain amount of discord).   
For Gaussian states, yet, it is conjectured to be optimal, i.e. ${D}_G \ug {\cal D}$~\cite{Adesso2010a, Giorda2010a, GIORDA,OLIVARES}.
Gaussian discord is analytically computable for all two-mode Gaussian states (notably, all such states, except product states, have non-zero Gaussian discord). 

The correlation matrix (see Sec. \ref{time-harm}) can be arranged in a $(2\times2)$-block form as
\begin{equation}
{C} = \left( \begin{array}{cc} {C}_1 & {C}_3 \\ {C}_3^\top & {C}_2 \end{array} \right).
\end{equation}
From the correlation matrix ${C}$, five symplectic invariants \cite{Ferraro2005a} can be constructed
\begin{eqnarray}
	&I_1=4 \;\mbox{Det}[{C}_1], \;\;\; I_2=4\; \mbox{Det}[{C}_2], \;\;\; I_3=4 \; \mbox{Det}[{C}_3],&
	\nonumber \\
	& I_4=16\;  \mbox{Det}[{C}], \;\;\; I_\Delta=I_1 + I_2 + 2I_3, \nonumber&
\end{eqnarray}
and two symplectic eigenvalues
\begin{gather}
	\lambda_{\pm}=\sqrt{\frac{I_\Delta \pm \sqrt{I_\Delta^2-4I_4}}{2}}.
	\label{EqSympEig}
\end{gather}
Gaussian discord can be defined in terms of these quantities (which are invariant under local unitary operations) as
\begin{equation}
	{D}_G(B|A) = f(\sqrt{I_1})-f(\lambda_{-})-f(\lambda_{+})+f(\sqrt{W}),
\end{equation}
where
{\begin{equation}
	f(x) \equiv \left( \tfrac{x+1}{2} \right) \log_2 \left( \tfrac{x+1}{2} \right) - \left( \tfrac{x-1}{2} \right) \log_2 \left( \tfrac{x-1}{2} \right)
\end{equation}}
and 
\begin{widetext}
\begin{gather}
	W = \begin{cases}
		\dfrac{2I_3^2+(I_1-1)(I_4-I_2)+2|I_3|\sqrt{I_3^2+(I_1-1)(I_4-I_2)}}{(I_1-1)^2} & \;\;\; \text{if } (I_4 - I_2 I_1)^2 \leq (1+I_1)I_3^2(I_2+I_4)\\\\
		\dfrac{I_2 I_1 - I_3^2 + I_4 - \sqrt{I_3^4+(I_4-I_2 I_1)^2-2I_3^2(I_4+I_2 I_1)}}{2I_1} & \;\;\; \text{otherwise},
	\end{cases}
\end{gather}
\end{widetext}
The analogous quantity ${D}_G(A|B)$ can be computed by exchanging $I_1$ with $I_2$ in the above formulas and describes the correlations retrieved by measuring subsystem $B$ first (instead of subsystem $A$). For the initial states considered in Section \ref{HO}, exchanging the role of the two subsystems has no effect, so that the two quantities coincide and we simply call them $D_G$.

\subsubsection{Qubits}

For a two-qubit system, the local measurement on system A is written as $\Pi_l^A (\theta,\phi)=\ket{l}_A\bra{l}\otimes\openone_B\;(l =1,2)$ 
with 
\begin{eqnarray}
\ket{1}&\ug&\cos\left(\frac{\theta}{2}\right)\ket{e}+e^{i\phi}\sin\left(\frac{\theta}{2}\right)\ket{g}\,,\\
\ket{2}&\ug&\sin\left(\frac{\theta}{2}\right)\ket{e}-e^{i\phi}\cos\left(\frac{\theta}{2}\right)\ket{g}
\end{eqnarray}
being orthogonal single-qubit states.
The total amount of classical correlations [\cf\eqref{c-cor}] reads 
\begin{equation}
\mathcal{C}(B|A){=}\max_{\theta,\phi} \left[S(\rho_{B}){-}\sum_l p_l S \left( \frac{{\rm Tr}_A[\Pi_l^A (\theta,\phi)\rho\Pi_l^A (\theta,\phi)]}{p_l} \right) \right].\nonumber
\end{equation}

\subsection{Entanglement}
\subsubsection{Harmonic oscillators}
{In Section \ref{HO}, we use {\it logarithmic negativity} for measuring entanglementof harmonic oscillators. It directly stems from the positive partial transpose (PPT) criterion \cite{Peres1996a} for discriminating entangled and separable states. A bipartite separable state can be written by definition as {$\rho_{SEP} = \sum_i p_i \rho_A^{(i)} \otimes \rho_B^{(i)}$, with 
$\rho_A^{(i)}$, $\rho_B^{(i)}$ being states of the subsystems $A$ and $B$ respectively and $p_i$ being probabilities.} It's easy to see that its partial transpose with respect to one subsystem (say A) $\rho_{SEP}^{\top_A} = \sum_i p_i \rho_A^{(i)\top_A} \otimes \rho_B^{(i)}$ is still a valid density matrix and hence is positive definite. Conversely, a non positive partial transpose always indicates the presence of entanglement. The logarithmic negativity quantifies how negative the partial transpose is.

For $1\otimes1$-modes gaussian states the PPT criterion is both necessary and sufficient \cite{Simon2000a}. This also implies that the logarithmic negativity is a faithful measure of entanglement. In terms of correlation matrix ${C}$, partial transposition is equivalent to changing the sign of momenta for a subsystem (say A). The partial transpose ${C}^{\top_A}$ is positive if and only if its symplectic eigenvalue $\tilde \lambda_{-}$ is greater than $1/2$ \cite{Ferraro2005a}. The symplectic eigenvalue $\tilde \lambda_{-}$ can be found, analogously to eq \eqref{EqSympEig}, as
\begin{gather}
	\tilde \lambda_{-}=\sqrt{\frac{\tilde I_\Delta - \sqrt{\tilde I_\Delta^2-4I_4}}{2}},
\end{gather}
where now $\tilde I_\Delta=I_1 + I_2 - 2I_3$ (note the change of sign due to partial transposition). The logarithmic negativity $E_\mathcal{N}$ is then defined as
\begin{equation}
	E_\mathcal{N} = \max\{ 0, -\log(2 \tilde\lambda_{-}) \}.
\end{equation}
Consistently $E_\mathcal{N} > 0$ when $\tilde\lambda_{-} < 1/2$.
}
\subsubsection{Qubits}

The {\it concurrence} is a measure of entanglement of two-qubit states, which is given by
\begin{equation}
C(\rho) =\max(\lambda_1-\lambda_2-\lambda_3-\lambda_4,0)\,,\label{Conc}
\end{equation}
where $\{\lambda_i\}$ are the square roots of the eigenvalues of matrix $M(\rho){=}\rho (\hat\sigma_{1y} \hat\sigma_{2y}) \rho^*(\hat\sigma_{1y} \hat\sigma_{2y})$
sorted in decreasing order while $\rho^*$ is the complex conjugate of density matrix $\rho$.
For two-qubit $X$ states
\begin{equation}
\rho=\left(\begin{matrix}
a & 0 & 0 & w^* \\
0 & b & z^* & 0 \\
0 & z & c & 0 \\
w & 0 & 0 & d \\
\end{matrix}\right)
\end{equation}
\eq \eqref{Conc} in this case becomes
\begin{equation}
C(\rho)=\max\left[2(\vert w\vert-\sqrt{bc}),2(\vert z\vert-\sqrt{ad}),0\right]\,.
\end{equation}
For the initial states addressed in Section \ref{Qubits-in} we thus find
\begin{equation}
C[\rho(0)]\ug\max [\frac{1}{2}\left(\left|\rho_{23}(0)\right| +\xi_S^2-1\right),0]=0\,,
\end{equation}
where we have taken into account \eq\eqref{qubitdomain} in the main text.

\end{appendix}


\begin{thebibliography}{99}
\bibitem{footnote-cv} The former point is such that the variances $\langle (\hat{X}_1 \meno \hat X_2)^2 \rangle \ug 2[C_{11}(0) \meno C_{13}(0)]$ and $\langle (\hat Y_1 \piu \hat Y_2)^2 \rangle \ug 2[C_{11}(0) \piu C_{24}(0)]$ are both zero, while for the latter one the variances $\langle (\hat X_1 \piu \hat X_2)^2 \rangle \ug 2[C_{11}(0) \piu C_{13}(0)]$ and $\langle (\hat Y_1 \meno \hat Y_2)^2 \rangle \ug 2[C_{11}(0) \meno C_{24}(0)]$ simultaneously vanish.
\end{thebibliography}
\end{document}